\documentclass[%
showkeys,
amsmath,amssymb, aps, prd, nofootinbib
]{revtex4-1}

\usepackage{graphicx}
\usepackage{dcolumn}
\usepackage{multirow}
\usepackage{bm}

\newcommand{\be}{\begin{equation}}
\newcommand{\ee}{\end{equation}}
\newcommand{\pa}{\partial}
\newcommand{\nn}{\nonumber}

\begin{document}

\title{Inclusive and effective bulk viscosities in the hadron gas}
\author{J.-B.~Rose$^{1,2}$, J.~M.~Torres-Rincon$^{1}$, and H.~Elfner$^{3,1,2}$}
\affiliation{$^1$Institute for Theoretical Physics, Goethe University,
Max-von-Laue-Strasse 1, 60438 Frankfurt am Main, Germany}
\affiliation{$^2$Frankfurt Institute for Advanced Studies, Ruth-Moufang-Strasse 1, 60438
Frankfurt am Main, Germany}
\affiliation{$^3$GSI Helmholtzzentrum f\"ur Schwerionenforschung, Planckstr. 1, 64291
Darmstadt, Germany}
\date{\today}

\begin{abstract}
We estimate the temperature dependence of the bulk viscosity in a relativistic hadron gas. Employing the Green-Kubo formalism in the SMASH (Simulating Many Accelerated Strongly-interacting Hadrons) transport approach, we study different hadronic systems in increasing order of complexity. We analyze the (in)validity of the single exponential relaxation ansatz for the bulk-channel correlation function and the strong influence of the resonances and their lifetimes. We discuss the difference between the inclusive bulk viscosity of an equilibrated, long-lived system, and the effective bulk viscosity of a short-lived mixture like the hadronic phase of relativistic heavy-ion collisions, where the processes whose inverse relaxation rate are larger than the fireball duration are excluded from the analysis. This clarifies the differences between previous approaches which computed the bulk viscosity including/excluding the very slow processes in the hadron gas. We compare our final results with previous hadron gas calculations and confirm a decreasing trend of the inclusive bulk viscosity over entropy density as temperature increases, whereas the effective bulk viscosity to entropy ratio, while being lower than the inclusive one, shows no strong dependence to temperature.
\end{abstract}

\maketitle

\section{\label{intro}Introduction}

Transport coefficients give insights about the microscopic dynamics of interacting matter close to equilibrium. The shear viscosity over entropy density $\eta/s$ is the most extensively studied transport coefficient in relativistic heavy-ion collisions (RHICs); since the first viscous hydrodynamic calculations became available in 2008~\cite{Song:2007fn,Luzum:2008cw}, the extraction of $\eta/s$ and its temperature dependence has been increasingly refined over the last decade. However, the situation is a bit different for the case of the bulk viscosity. Since AdS/CFT calculations (as models for QCD dynamics in the strong coupling) imply that it is very small for nearly-conformal systems~\cite{Kovtun:2004de}, the bulk viscosity $\zeta$ (and its corresponding dimensionless ratio $\zeta/s$)---which can be thought of as the resistance to uniform expansion/compression of a fluid---has not been subject to the same extended treatment as $\eta$ in the context of RHICs~\cite{Gale:2013da}. It should be pointed out that although $\zeta$ is identically zero in conformal fluids~\cite{Weinberg:1972kfs}, and that QCD approaches conformality in the limit of high energies/temperatures~\cite{Borsanyi:2013bia,Bazavov:2017dsy}, there is no evidence that the nuclear matter which is produced in accelerators (even at the highest LHC energies) is a conformal fluid. Moreover, the system becomes less and less scale invariant as the system cools down with time~\cite{Karsch:2007jc}. 

Although not exhaustive, some studies on the effect of bulk viscosity on some observables such as elliptic flow~\cite{Denicol:2009pe,Denicol:2009am} and particle spectra~\cite{Monnai:2009ad} were made. More recently, the bulk viscosity has started attracting more attention since it was pointed out by phenomenological studies in hybrid models that the inclusion of bulk viscosity as described by~\cite{Denicol:2014vaa} was important in some cases to properly reproduce simultaneously the radial and azimuthal flow anisotropies~\cite{Ryu:2015vwa,Ryu:2017qzn}. Most notably, the first quantitative extractions of shear and bulk viscosities employing Bayesian techniques have recently appeared~\cite{Bernhard:2016tnd, Bernhard:2019bmu}. In these works, the functional form of the temperature dependence of the transport coefficients influences the prior and therefore an external input for these is very important. In particular, the bulk viscosity is expected to have a peak around the transition from hadronic matter to the quark-gluon plasma~\cite{Bernhard:2019bmu}. Close to the phase transition at vanishing baryochemical potential calculations based on lattice QCD indicate an enhancement of the bulk viscosity~\cite{Karsch:2007jc}. Above the crossover temperature, a fast drop-off is also suggested by quasi-particle models~\cite{Paech:2006st,Bluhm:2010qf}. 

On the purely hadronic side, theoretical calculations of the bulk viscosity are notoriously more complicated than those of the shear viscosity, and as such are scarcer. However, using different models and computational techniques, the temperature dependence of the bulk viscosity of a hadron gas was presented e.g. in Refs.~\cite{gavin1985transport,Davesne:1995ms,Chen:2007kx,NoronhaHostler:2008ju,FernandezFraile:2008vu,Chakraborty:2010fr,Khvorostukhin:2010aj,Moroz:2011vn,Lu:2011df,Dobado:2011qu,Dobado:2012zf,Ozvenchuk:2012kh,Moroz:2013vd,Kadam:2015xsa,Ghosh:2015mda,Attems:2016ugt,Rougemont:2017tlu}. Results from the various calculations differ from one another by an order of magnitude or more, as we will see when comparing our own results with some of these calculations.

Among these calculations we will pay special attention to those restricted to very low temperatures where pions dominate the hadronic mixture. In this regime the interactions of pions can be described by chiral perturbation theory (and its unitarized version to describe the resonant energy domain). One of these calculations~\cite{FernandezFraile:2008vu} applied a diagrammatic Green-Kubo method and predicted a double bump structure for $\zeta$ at low $T$. The first of these bumps was explained from the explicit conformal breaking due to the pion mass, while the second was related to the conformal anomaly appearing at temperatures close to the crossover. A calculation with similar interactions but using the Boltzmann-Uehling-Uhlenbeck equation~\cite{Dobado:2011qu} further commented that the addition of a pion pseudochemical potential was also necessary for a consistent treatment of pion elastic collisions. However, in Ref.~\cite{Lu:2011df} the focus was on the much slower $2\leftrightarrow4$ pion inelastic processes and obtained a very different value of the bulk viscosity (diverging at $T=0$). These rather different calculations illustrate the effect of including or excluding particle number-changing processes: whereas Ref.~\cite{Lu:2011df} uses the idea that the slowest processes (inelastic collisions) should dominate the value of $\zeta$, Refs.~\cite{FernandezFraile:2008vu,Dobado:2011qu} argue that such processes are so slow that they cannot be effective at all in RHICs. In this paper we will clarify the conceptual difference between the two points of view---distinguishing between ``inclusive'' and ``effective'' bulk viscosities---by addressing this coefficient using the microscopic simulation code SMASH (Simulating Many Accelerated Strongly-interacting Hadrons).

In the following we will present various results for the bulk viscosity in simple hadronic systems of various chemical compositions, amongst which hadronic predictions for hydrodynamical calculations of RHICs. Some more technical considerations that have to be taken into account in order to obtain them will also be discussed.

In Sec.~\ref{sec:calibration} we introduce the methodology to extract the bulk viscosity via a Green-Kubo relation and introduce the SMASH transport approach. In Sec.~\ref{sec:bulk_pion} we apply the model to a simple relativistic gas interacting with constant cross section, where comparison with the corresponding Chapman-Enskog solution will calibrate our model in terms of systematic uncertainties. In Sec.~\ref{sec:full_bulk_section} we show that adding resonances to the system requires a revisiting of the assumption made for the form of the correlation function. We show how the simple exponential decay ansatz breaks down, and further analyze the effect of the resonance lifetimes. In Sec.~\ref{sec:hadrongas} we apply the method to the full hadron gas for several temperatures and box sizes. We introduce definitions for the inclusive and effective bulk viscosities and present final results for both $\zeta/s$ and $\zeta_{\rm{eff}}/s$, comparing with previous calculations. Finally, in Sec.~\ref{sec:conclusions} we summarize our work.

\section{Methodology~\label{sec:calibration}}

\subsection{Green-Kubo formalism}

In this work we apply the Green-Kubo formalism~\cite{green1952markoff,green1954markoff,kubo1957statistical} to obtain the bulk viscosity coefficient of different systems. While different versions of the Green-Kubo formula exist in the literature depending on the system and thermodynamical ensemble used, the most general form reads~\cite{mori1962collective,luttinger1964theory,Zwanzig,Zubarev,Horsley:1985dz}
\be \label{eq:corre}
\zeta = \frac{V}{T} \int_{0}^{\infty} dt \ \langle \Delta \Pi(0) \Delta \Pi(t) \rangle \ ,
\ee
where $V$ is the volume of the system, $T$ is the temperature, and $\Delta \Pi (t) \equiv \Pi (t) - \langle \Pi \rangle$ is a fluctuation around the thermodynamical equilibrium average. The variable $\Pi$ is defined as
\be \label{eq:Pi}
\Pi(t) \equiv P (t) -  \left( \frac{\partial P}{\partial \epsilon} \right)_n   \epsilon (t) - \left( \frac{\partial P}{\partial n} \right)_\epsilon n (t) \ ,
\ee
where $P(t)=\frac13 T^{ii} (t)$~\footnote{Our Minkowski metric convention is mostly minus $g^{\mu\nu}=(+,-,-,-)$ .} is the (instantaneous) pressure, $\epsilon(t)=T^{00}(t)$ the (instantaneous) energy density and $n(t)=j^0(t)$ the (instantaneous) particle density. All components of the energy-momentum tensor $T^{\mu\nu}$ and the particle 4-current $j^\nu$ are understood to be averaged over $V$,
\be T^{\mu \nu}(t)=\frac{1}{V} \int d{\bf r} \  T^{\mu \nu} (t,{\bf r}) \ , \quad 
 j^\mu(t)=\frac{1}{V} \int d{\bf r} \  j^\mu (t,{\bf r}) \ .
\ee

In $\Pi(t)$ appear two thermodynamical quantities: the speed of sound at constant number density and the compressibility at constant energy density. These quantities naturally appear in the source function (left-hand side) of the Boltzmann equation when considering the bulk viscosity of a gas with a conserved (net) particle number~\cite{Hosoya:1983xm,gavin1985transport,Sasaki:2008fg,Torres-Rincon:2012sda}.

For later reference the adiabatic speed of sound at constant entropy per particle $S=s/n$ is related to these two quantities as~\cite{Hosoya:1983xm,Torres-Rincon:2012sda}
\be v_S^2=\left( \frac{\partial P}{\partial \epsilon} \right)_{s/n} =  \left( \frac{\partial P}{\partial \epsilon} \right)_n + \frac{n}{w} \left( \frac{\partial P}{\partial n} \right)_\epsilon \ , \label{eq:vs} \ee
where $w=\epsilon+P$ is the enthalpy density. Expressions for all these quantities as functions of temperature are given in App.~\ref{app:correl}.

For convenience let us define the autocorrelation function 
\be C_\zeta (t) \equiv \langle \Delta \Pi(0) \Delta \Pi(t) \rangle \ , \ee
which will be extracted from our numerical SMASH simulations and integrated over time as in Eq.~(\ref{eq:corre}). For other transport coefficients such as the shear viscosity or electric conductivity in dilute systems~\cite{Demir:2008tr,Plumari:2012ep,Wesp:2011yy,Rose:2017bjz,Hammelmann:2018ath,Rose:2020sjv} it is generally assumed that the correlation function takes the form of a decaying exponential. This ansatz can be motivated by the relaxation-time approximation of the Boltzmann equation~\cite{Bhatnagar:1954zz} or by the causal hydrodynamic equations~\cite{Muronga:2003ta}. This particular form should always be confirmed {\it a posteriori} within the precision of the data acquired. We have checked it for every case presented in this work. If this ansatz is inadequate, then it will be not used (as will happen for hadron mixtures later on). For the bulk viscosity, this ansatz reads
\begin{equation}
  C_\zeta (t) = C_\zeta (0) \ e^{-t/\tau_\zeta} \ ,
  \label{eq:correl_ansatz}
\end{equation}
where $\tau_\zeta$ is the bulk relaxation time of the system. From Eq.~(\ref{eq:corre}), it follows that
\begin{equation}
  \zeta = \frac{C_\zeta (0) V \tau_\zeta}{T} \ .
  \label{final_bulk_eq}
\end{equation}

In some previous works the relaxation time $\tau_\zeta$ has been estimated to be related to the mean free time of the particles, i.e. the average time between collisions used for other transport coefficients as well. However this introduces a new source of uncertainty, as different transport coefficients are sensitive to different transport mechanisms. For example, while the mean free path is inversely proportional to the total cross section, the shear viscosity is sensitive to the ``transport cross section'', which could be a factor of 2 smaller than the total cross section in a $p-$wave scattering~\cite{Rose:2017bjz}. More importantly, the use of the mean free time misses the dependence of $\tau_\zeta$ on the resonance lifetimes, which was noted to be very important in the shear viscosity case~\cite{Rose:2017bjz}. As we will see, this will also prove to be particularly significant for the bulk viscosity, where the relaxation times for elastic and inelastic processes are, in general, completely different~\cite{Jeon:1995zm,Lu:2011df}.

It is helpful to realize that the value of $C_\zeta(0)$ is an equilibrium quantity. From its definition,
\begin{align} 
C_\zeta(0) & = \left\langle  \int \frac{d^3p_1}{(2\pi)^3} \ [f_1 (0)-f^{{\rm eq}}_1] \left[  \frac{p_1^2}{3 E_{1}} -  \left( \frac{\partial P}{\partial \epsilon} \right)_n E_{1} - \left( \frac{\partial P}{\partial n} \right)_\epsilon  \right] \right. \nn \\ 
& \times \, \, \, \,    \left.  \int \frac{d^3p_2}{(2\pi)^3} \ [f_2 (0)-f^{{\rm eq}}_2] \left[  \frac{p_2^2}{3 E_{2}} -  \left( \frac{\partial P}{\partial \epsilon} \right)_n E_{2} - \left( \frac{\partial P}{\partial n} \right)_\epsilon  \right] \right\rangle \ ,
\end{align}
where $f_1^{{\rm eq}}=f^{{\rm eq}}({\bf p}_1)$ is the (spatially-averaged) distribution function in equilibrium e.g. the Maxwell-Boltzmann function $f^{{\rm eq}}_1=g\exp [-(E_1-\mu)/T]$ ($g$ is the internal degeneracy of the particle), and $E_{1}=\sqrt{p_1^2+m^2}$.

To compute $C_\zeta(0)$ we need to know the equal-time correlation function of the spatially-averaged fluctuation of the distribution function, 
\be \delta f_1 (t) \equiv f_1(t)- f_1^{{\rm eq}} \ , \ee
for which we can directly apply the result of~\cite{lifschitz1983physical} for the 2-point correlation function,
\be \label{eq:fluct} \langle \delta f_1(0) \delta f_2(0) \rangle = \frac{(2\pi)^3}{V} f^{{\rm eq}}_1 \delta^{(3)} (p_1-p_2) \ . \ee
Combining these, we obtain
\be \label{eq:C0}
C_\zeta(0) V= \int \frac{d^3p}{(2\pi)^3} f^{{\rm eq}}(p) \frac{1}{E_p^2} \left[ \frac{p^2}{3} -  \left( \frac{\partial P}{\partial \epsilon} \right)_n E_p^2 -  \left( \frac{\partial P}{\partial n} \right)_\epsilon E_p \right]^2 \ . 
\ee

Incidentally, this formula exactly coincides with the quantity $T\zeta/\tau^\zeta_R$ derived in~\cite{Torres-Rincon:2012sda} for a (Bose) gas with binary interactions. The result in~\cite{Torres-Rincon:2012sda} uses the relaxation time approximation, where one identifies $\tau^\zeta_R \simeq \tau_\zeta$. Using the expressions given in App.~\ref{app:correl} for the different thermodynamics quantities appearing in (\ref{eq:C0}), one can compute the explicit temperature dependence of $C_\zeta(0)$.

\subsection{Hadron gas modeling: SMASH}

In this work we use the SMASH transport approach~\cite{Weil:2016zrk,dmytro_oliinychenko_2019_3485108} to simulate infinite hadronic matter in a box with periodic boundary conditions. In SMASH, all well-established hadrons of the PDG~\cite{Tanabashi:2018oca} are included, with their interactions modeled by resonance excitation and decay, elastic as well as inelastic $2 \leftrightarrow 2$ processes. 

At this point it is important to mention that the $V$ used in our analysis is the entire simulation box volume, instead of a subvolume of the whole system. By doing so, we get that the total energy and total particle number are conserved (at least in simple systems), bringing the system in a sort of microcanonical ensemble over $V$. Therefore $\Delta n(t)=\Delta \epsilon(t)=0$ , and the correlation function reduces to
\be \label{eq:Ctsimp} C_\zeta (t)=\langle \Delta P(0) \Delta P(t) \rangle \ . \ee

The instantaneous pressure $P(t)$ is extracted from the energy-momentum tensor $T^{\mu\nu} (t)$ of the equilibrated system, following the methodology described in~\cite{Rose:2017bjz,Hammelmann:2018ath}. Such simulations provide the complete phase-space information of all particles in the system, which are in this case discrete, and given at specific time steps. For this situation, we can define the components of the energy-momentum tensor as
\be
T^{\mu\nu} (t) = \frac{1}{V} \sum_{i=1}^{N} \frac{p^\mu_i (t) p^\nu_i (t)}{p^0_i (t)} \ , \label{tmunu_discrete}
\ee
where $N$ is the total number of particles in $V$, $p^\mu_i$ is a component of the momentum 4-vector associated with particle $i$.

The averaging contained in the correlation function~(\ref{eq:Ctsimp}) also has to be defined for the discrete times $t \equiv u\Delta t$ at which the information is available,
\begin{align}
C_\zeta(t) =& \langle \Delta P(0) \Delta P(u \Delta t) = \lim_{K\to\infty}\frac{1}{K-u} \sum_{s=0}^{K-u} \Delta P(s\Delta t) \Delta P(s\Delta t + u\Delta t) \ ,
\label{eq:correlator_discrete}
\end{align}
where $K$ is the total number of considered time steps, $u$ is a positive integer with $u < K$ and $\Delta t$ is the time interval between each time step. It is numerically challenging to take the limit of $K \to \infty$ in Eq.~(\ref{eq:correlator_discrete}) and thus the relative error of any numerical computation of the correlation function necessarily increases rather quickly with time and eventually reaches a state of pure noise, as one can see for example on Fig.~\ref{fig:bulk_correls_pion}.

\section{Simple gas with elastic interaction~\label{sec:bulk_pion}}

A single-component relativistic gas interacting through elastic collisions provides the first example to test our method. In the case of a gas with constant, isotropic cross section (hard-sphere gas) the bulk viscosity is zero in the nonrelativistic and the ultrarelativistic limits~\cite{Weinberg:1972kfs,Karsch:2007jc}. However in an intermediate domain of temperatures the bulk viscosity is small, but nonzero. Without loss of generality we will assign a mass to the particles $m=138$ MeV, and internal degeneracy of $g=3$ (resembling a pion gas but interacting with a constant cross section of $\sigma=20$ mb).

Such a hard-sphere gas has been studied before in the context of the bulk viscosity. Its value has been extracted analytically e.g. in~\cite{AndersonKox} by linearizing the collision term of the Boltzmann equation using the Chapman-Enskog approximation to first order (see also~\cite{Moroz:2013vd}). More generally, by modifying the numerical codes used in~\cite{Dobado:2011qu,Torres-Rincon:2012sda} we can extend the Chapman-Enskog expansion for this system to higher orders to check convergence. This will help us to calibrate the Green-Kubo calculation in this simple case.

\begin{figure}[ht]
  \centering
  \includegraphics[scale=0.5]{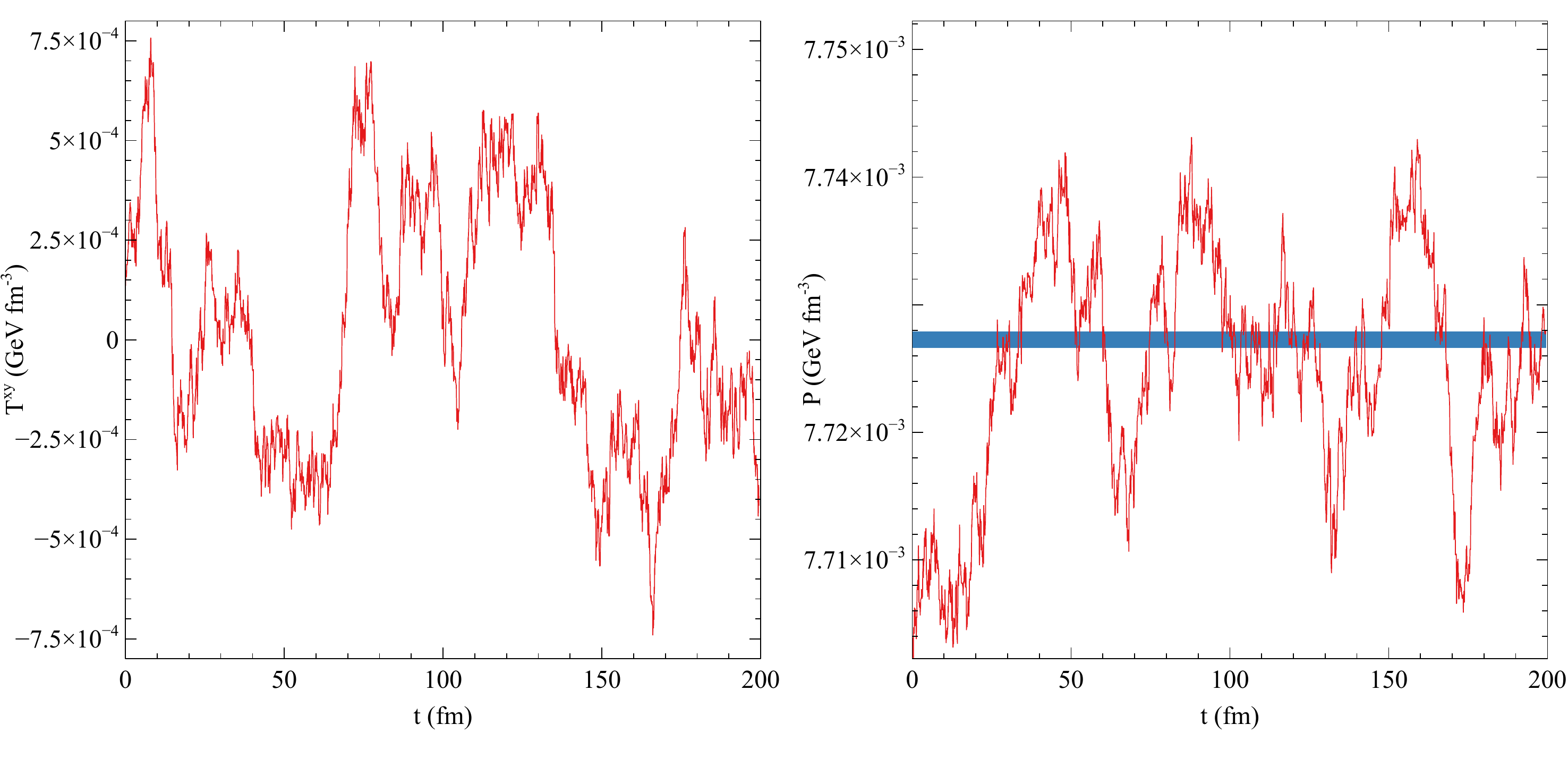}
  \caption{Sample fluctuations for the shear (left) and bulk (right) channels for a gas of particles with $m=138$ MeV interacting via a constant $\sigma = 20$ mb cross-section at a temperature of 125 MeV in a $V=(20$ fm$)^3$ volume. The thick blue band on the right panel is the average pressure, with its uncertainty.  }
  \label{fig:sample_fluct_shear_bulk}
\end{figure}

To start with, it is instructive to look at a sample of the measured fluctuations of the pressure in such a system and to compare it to the off-diagonal energy-momentum fluctuations $T^{xy}(t)$ associated with shear viscosity (see~\cite{Rose:2017bjz,Rose:2017ntg}), which is done on Fig.~\ref{fig:sample_fluct_shear_bulk}. While the fluctuating nature of the two signals appears relatively similar at first glance, there are significant differences between them. First, the amplitude of the signal for $T^{xy}(t)$ is roughly 25 times larger than the case of the pressure (notice the different $OY$ axis scales). Second is the fact that pressure does not oscillate around zero, and thus an average pressure needs to be subtracted in the correlation function. This is not as trivial as one could think, as the average pressure also introduces a statistical error which can be non-negligible. While the calculation of the correlation function is done over 4000 time steps spanning 200 fm as in the case of the shear viscosity (see~\cite{Rose:2017bjz}), we find that in order to get results in which the statistical error does not completely wash out the signal, the averaging of the pressure requires much larger data sets. We determined that for the studied cases, an averaging going over 100 000 time steps spanning 5000 fm was sufficient, in the middle of which we perform the previously mentioned calculation of the correlation function. 

Note that, in principle, the thermodynamic pressure can be calculated analytically for such a gas assuming Boltzmann thermodynamics (e.g. via the ${\cal J}_{n,k}$ functions defined in App.~\ref{app:correl}). However, in more complex systems, although the SMASH equilibrium is very close to the grand canonical one, it can deviate from it slightly. For these bulk viscosity calculations, even minimal deviations in the average pressure of the order of a fraction of a percentage point can make a significant difference in the final signal, and thus such an analytical calculation would be highly non-trivial to perform to the required precision. Therefore, to keep our methodology consistent with the following sections, we always use the numerical extraction of the average pressure, instead of the Boltzmann expression. 

\begin{figure}[ht]
  \centering
  \includegraphics[width=76mm]{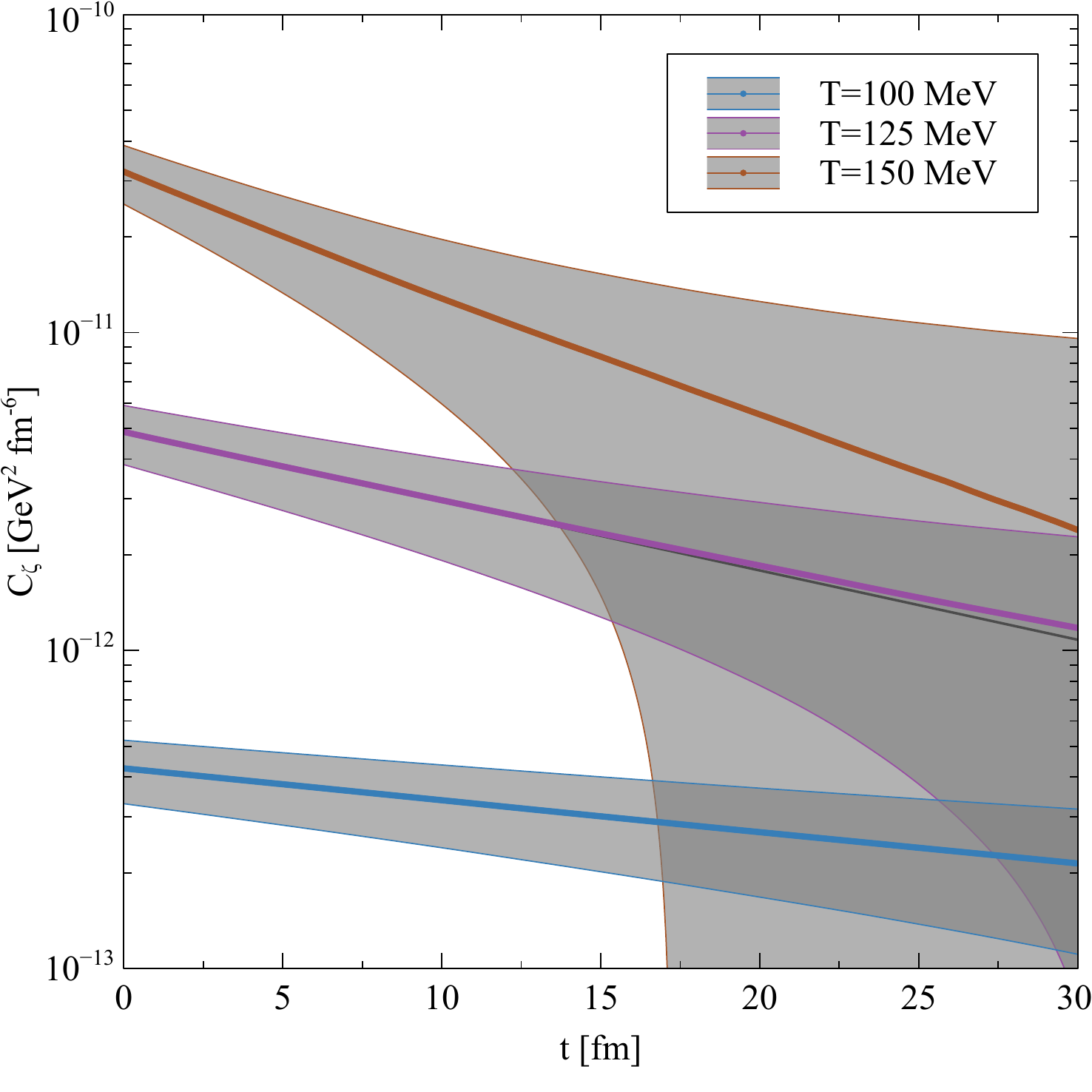}
  \includegraphics[width=76mm]{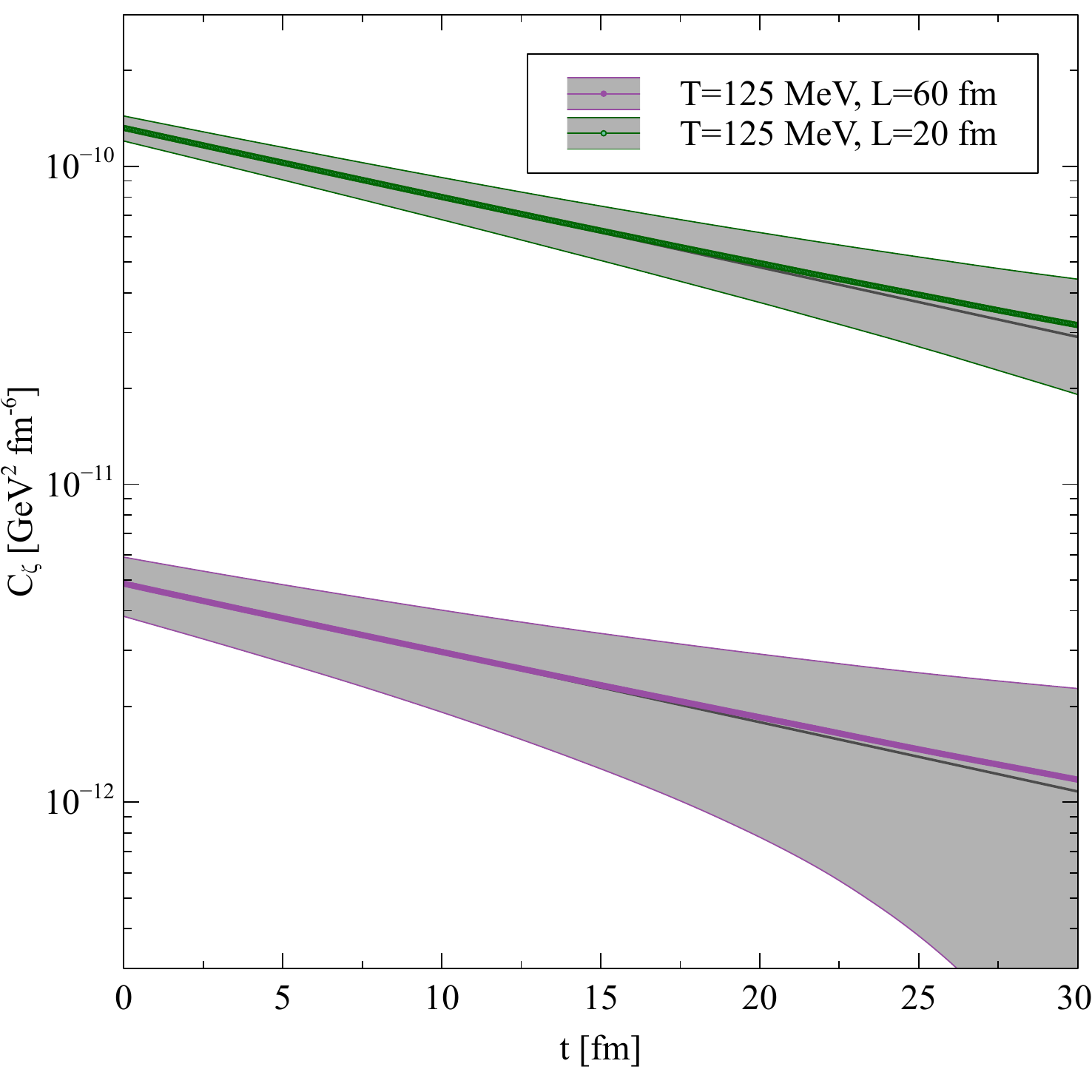}
  \caption{Bulk correlation function for a gas of $m=138$ MeV particles interacting through constant cross-section $\sigma = 20$ mb at various temperatures (left) and system volume (right), as well as exponential fits, fitting from $t=0-10$ fm.}
  \label{fig:bulk_correls_pion}
\end{figure}

Figure~\ref{fig:bulk_correls_pion} shows a collection of correlation functions. The left panel illustrates how a rising temperature leads to a steeper $C_\zeta(t)$ (which translates to a shorter relaxation time) as well as the expected increase of the statistical error as time increases. What is quite unique to the case of the bulk viscosity is that the initial value $C_\zeta (0)$ can have a relatively large statistical error, up to 20\% in this case, whereas in previous works the same error on the shear viscosity~\cite{Rose:2017bjz} or electrical conductivity~\cite{Hammelmann:2018ath} correlation function initial value was never larger than $\sim$ 6\%, which would barely be visible.

The right panel of Fig.~\ref{fig:bulk_correls_pion} additionally shows that the correlation function scales as the inverse of the volume of the box used [cf. Eq.~(\ref{eq:C0})], so that the increase of factor of 3 in the size of the box is reflected by a decreasing of a factor of 27 in $C_\zeta(t)$, with the slope being the same in both curves. Reducing the size of the box also reduces the relative statistical error, as the size of the fluctuations with respect to the average pressure diminishes as volume increases~(\ref{eq:fluct}). 

To calculate $\zeta$, one then has to strike a balance between having a system which is large enough for thermodynamic quantities to be calculated, but small enough that the signal does not get washed out by the statistical error. This volume might differ for each value of the temperature within the same system, as can be seen in Table~\ref{tab:volumes}, where we provide the specific volumes used for each temperature. Notice that for the ``smaller box'' set we have not applied smaller volumes than (20 fm)$^3$ because the number of pions inside the box becomes very scarce.

\begin{table}[ht]
\begin{center}
 \begin{tabular}{|c||c|c|c|c|c|} 
 \hline
 & $T=75$ MeV & $T=100$ MeV & $T=125$ MeV & $T=150$ MeV & $T=175$ MeV  \\ 
 \hline
 larger box, length & 200 fm & 100 fm & 60 fm & 40 fm & 30 fm\\ 
 smaller box, length & 60 fm  & 20 fm  & 20 fm & 20 fm & 20 fm \\
\hline
\end{tabular}
\caption{System size used for the single gas calculation of the bulk viscosity. For each temperature we use two sets of box sizes, denoted as ``larger box'' and ``smaller box'' in the figures. The volume of each box is $V={\rm length}^3$.
}
\label{tab:volumes}
\end{center}
\end{table}

\begin{figure}[ht]
  \centering
  \includegraphics[scale=0.4]{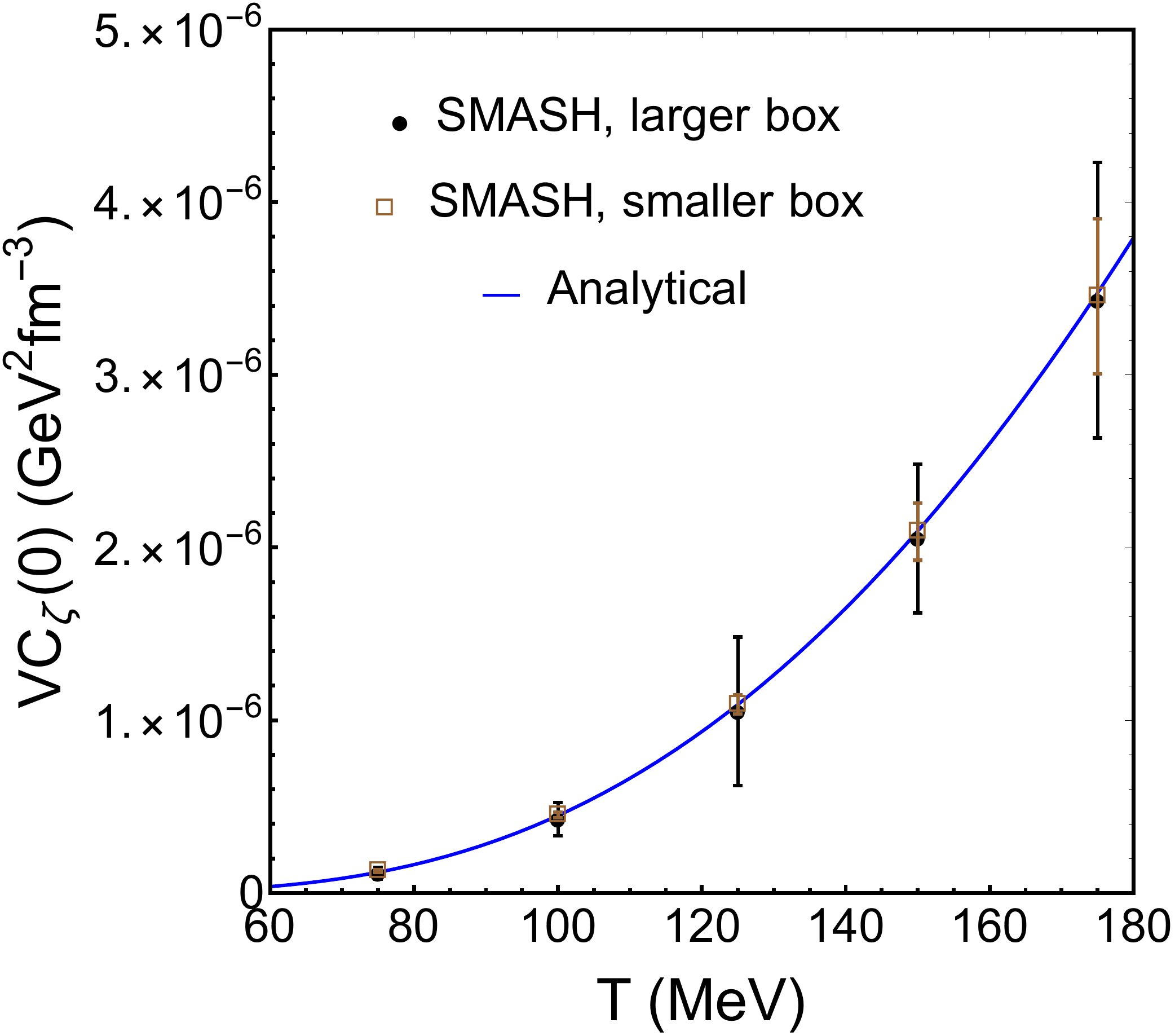}
  \caption{Correlation function at $t=0$ for a gas of $m=138$ MeV particles interacting through constant cross-section $\sigma = 20$ mb at various temperatures. The dots are the extracted results from SMASH, and the solid line is the analytical result from Eq.~(\ref{eq:C0}).}
  \label{fig:bulk_C0_pion}
\end{figure}

In Fig.~\ref{fig:bulk_C0_pion} we compare the $VC_\zeta(0)$ values for the simple gas as a function of the temperature. The symbols are the extracted values from our simulations using SMASH, including statistical errors. The solid line is the result of Eq.~(\ref{eq:C0}) at the corresponding temperature. Larger/smaller boxes refer to the values presented in Table~\ref{tab:volumes}; notice how the bigger volumes provide larger error bars, which is consistent with the larger error on the correlation function for these volumes. We observe that a very good agreement is obtained between the two, providing a nontrivial check on the method. 

We proceed to fit the correlation function to the exponential decay form \eqref{eq:correl_ansatz}. Notice that the relatively large uncertainty in $C_\zeta(t)$ makes it difficult to systematically decide where to stop the fit;  we will simply stop it at $t=5$ fm for this simple gas case.

\begin{figure}[ht]
\centering
\includegraphics[width=76mm]{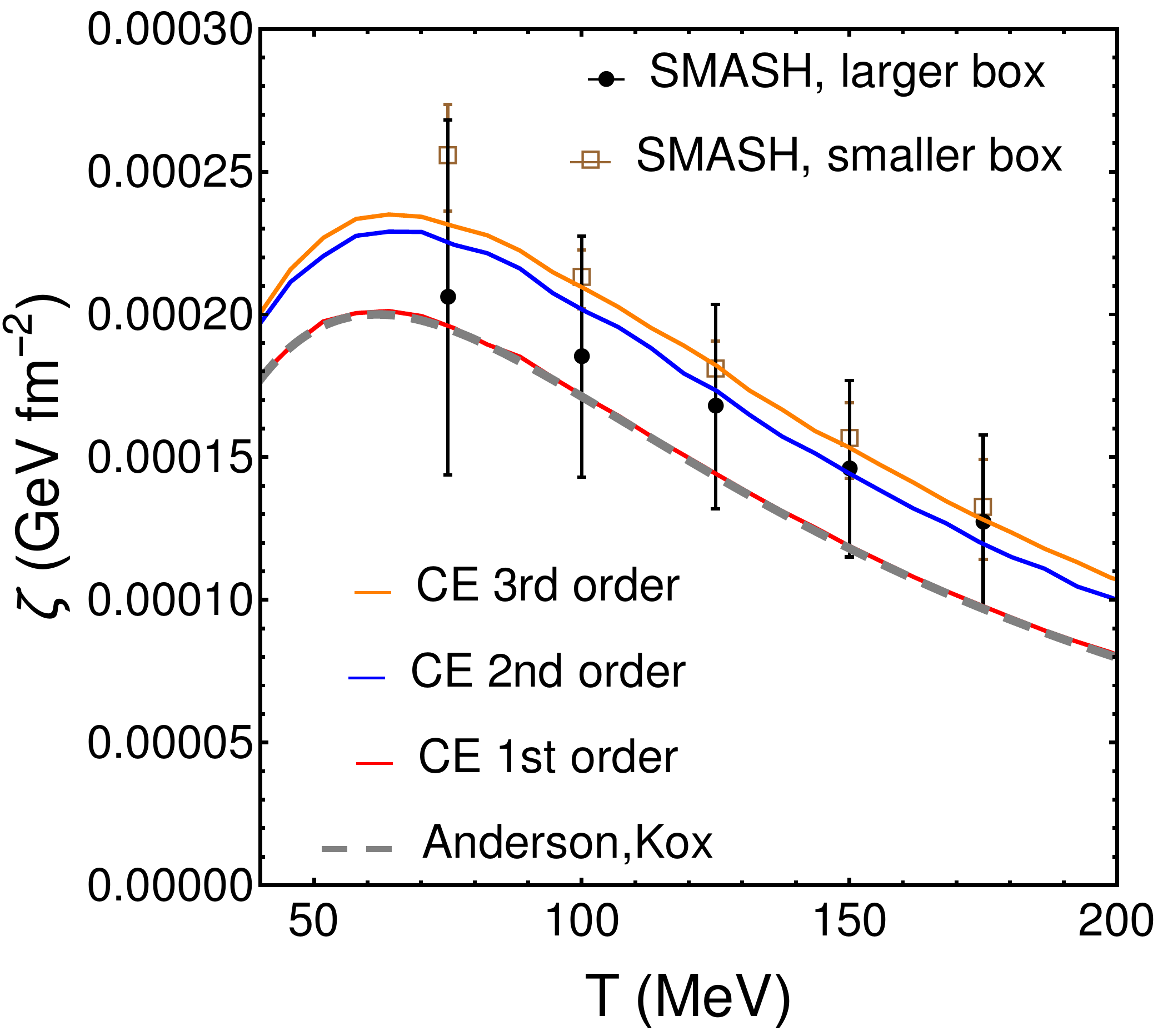}
\includegraphics[width=70mm]{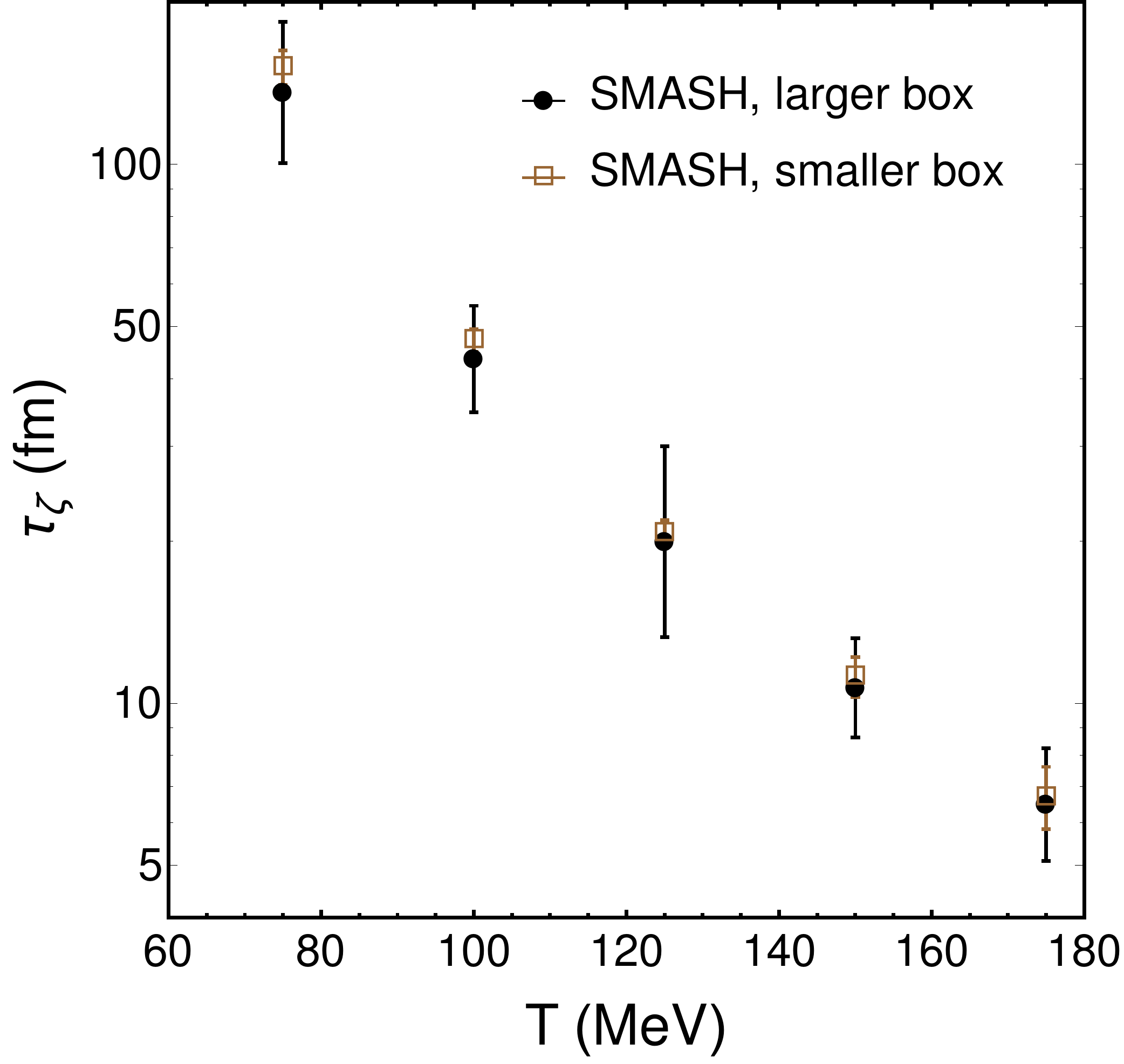}
\caption{Left panel: Bulk viscosity of a gas interacting through constant $\sigma = 20$ mb, as computed with the Green-Kubo formalism and compared to the third order Chapman-Enskog calculation. Dots are the result from SMASH for 2 different box sizes, ``smaller'' and  ``larger'' detailed in Table~\ref{tab:volumes}. The theoretical calculation comes from the Chapman-Enskog estimates in Refs.~\cite{AndersonKox} (1st order) and ~\cite{Torres-Rincon:2012sda} (first to third order). Right panel: Bulk relaxation time for the same system as a function of the temperature, for the same box sizes.}
\label{fig:bulk_pion}
\end{figure}

The bulk viscosity of this system calculated using the Green-Kubo formalism is compared to the semi-analytic Chapman-Enskog expansion in the left panel of Fig.~\ref{fig:bulk_pion} as a function of the temperature. The first-order Chapman-Enskog result is taken from~\cite{AndersonKox} and numerically re-calculated with the method of~\cite{Torres-Rincon:2012sda}, which also allows to go to third-order Chapman-Enskog where convergence is achieved.  The agreement is rather good for temperatures between 100 and 175 MeV, even for smaller system sizes. At low temperatures, the agreement starts to break down, and, although not shown here, crumbles completely at even lower temperatures. At those low temperatures we observe that the correlation function is still exponential, but the uncertainties are large: the number of pions at these temperatures is so small, that statistics are very poor, making results at lower temperatures than shown unreliable. In parallel, using large volumes to increase statistics also washes out the signal; this can be seen in the black dots, which not only underestimate the analytic $\zeta$ but also see their error bars increase significantly. We are thus unfortunately not able to observe the nonrelativistic limit in which the bulk viscosity turns to zero at $T\rightarrow0$. In the right panel of Fig.~\ref{fig:bulk_pion} we show the relaxation time $\tau_\zeta$. The uncertainty of the large volumes is again much larger than the smaller volumes, although the central values are compatible with those of the smaller volumes. The two panels of the figure show a good agreement between different calculations, validating the method for more complex systems.

\section{Effect of resonances~\label{sec:full_bulk_section}}

It is known that the presence of internal dynamical degrees of freedom (rotational, vibrational) as well as inelastic collisions---allowing a redistribution of internal energy in a more efficient way---contribute critically to the bulk viscosity~\cite{Weinberg:1971mx,Jeon:1995zm}. The latter might happen via strongly number-changing processes like the $2 \pi \leftrightarrow 4 \pi$ considered in~\cite{Lu:2011df} or the $N\bar{N} \rightarrow 5\pi$ annihilation, but also due to the presence of continuous resonance decay and recombination. These processes made a notable difference already for the shear viscosity~\cite{Rose:2017ntg}, and their role is expected to be even more relevant due to the nature of the bulk viscosity coefficient.

We start this discussion by presenting the bulk correlation function for the full hadron gas with resonances.  While we relegate its full analysis to Sec.~\ref{sec:hadrongas}, it will first serve us as a motivation to consider a more general ansatz for $C_\zeta(t)$ in the presence of several hadron species and resonances.

\subsection{Breakdown of the single exponential ansatz} 

A solid baseline has been established for the calculation of the bulk viscosity at temperatures between 100 MeV and 175 MeV after using a simple pion gas with constant cross section. We directly proceed to the case of the full hadron gas as described by SMASH v1.6~\cite{dmytro_oliinychenko_2019_3485108}. As mentioned earlier, this gas includes not only elastic but also inelastic processes, be they binary inelastic $2 \leftrightarrow 2$ interactions or, most commonly, indirect resonant $2 \leftrightarrow 1 \leftrightarrow 2$ reactions where two particles will form a resonance of mass $m$, width $\Gamma (m)$ and a sampled lifetime averaging at $\tau_{{\rm life}} = 1/ \Gamma (m)$, after which it will decay into two new daughter particles which can or not be of the same species as the original ones (it is also possible for resonances to scatter and form larger resonances with other particles during their lifetime; see~\cite{Weil:2016zrk} for details). Note that in order to calculate the average pressure of this system to an appropriate degree of precision, we require simulations to provide at least 5000 fm of equilibrium data; this is extremely costly in terms of computational power, and as such limits the breadth of the exploration of the parameter space. 

While in this system the particle number is not formally conserved by the inelastic processes, we make the approximation that the contribution of the third term in Eq.~\eqref{eq:Pi} is small with respect to the pressure fluctuations. We checked that particle number fluctuations are of the same order of magnitude as pressure fluctuations, but the former are multiplied by a small $(\pa P/\pa n)_{\epsilon}$ (see Fig.~\ref{fig:thermo}), largely reducing their contribution.

\begin{figure}[th]
  \centering
  \includegraphics[width=80mm]{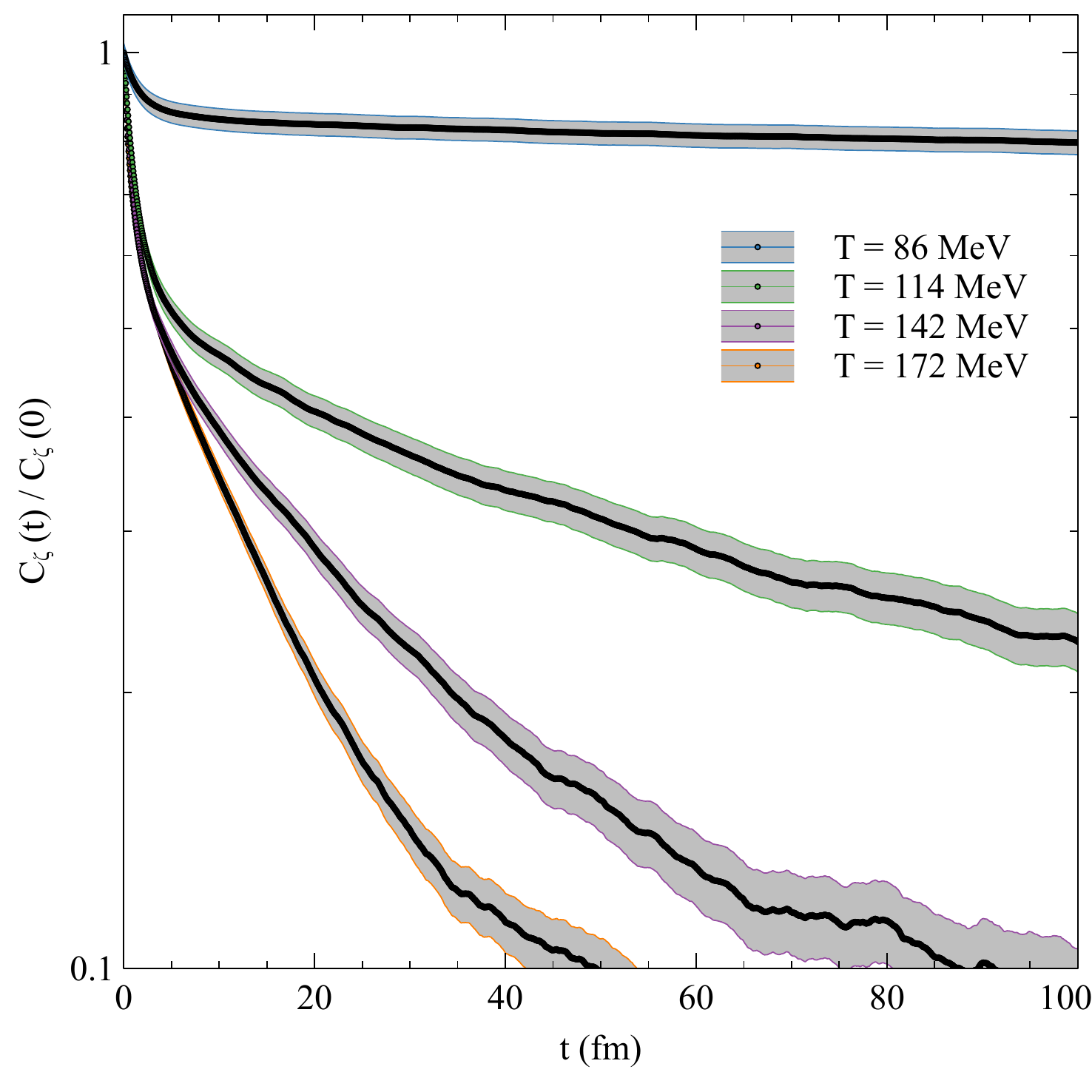}
  \caption{Bulk correlation functions for a full hadron gas, where every curve is normalized with its initial value.
  }
  \label{fig:bulk_correls_full}
\end{figure}

Let us consider the normalized (i.e. divided by their value at $t=0$) correlation functions presented on Fig.~\ref{fig:bulk_correls_full} for different temperatures. As is readily visible, these offer a considerably different picture as what we observed in the previous case in Fig.~\ref{fig:bulk_correls_pion}. First, the statistical errors are here much less significant than they previously were. This is expected, as the introduction of resonances (and thus of mass-changing processes which dissipates the otherwise purely kinetic energy) leads to a massive increase in the magnitude of the fluctuations with respect to the average pressure, and as such, it is expected that the error on the pressure plays a smaller role in this case.

More importantly, the correlation functions at all temperatures display a somewhat peculiar shape. In the first 2--3 fm a period of rapid exponential decorrelation is followed later on by a less abrupt decay over relatively long times before the relative error finally increases to a point where the signal is dominated by noise. It is evident that the correlation functions are not describable by a single exponential function, and one needs to abandon the simple ansatz in Eq.~(\ref{eq:correl_ansatz}).

\subsection{Single resonance gas: resonance lifetime and relaxation time}

To physically understand these important modifications of the shape of the correlation function, we look at a toy system with a minimal content of particles. Let us consider a box with pions (with their physical mass and isospin degeneracy) interacting through a single resonance, the $\rho$ meson. We switch off all other possible resonances and set to zero any contact cross section. This is a relatively simple system, in which we scale the lifetime of the decay $\rho \rightarrow \pi+\pi$ by a multiplicative factor. 

\begin{figure}[th]
  \centering
  \includegraphics[width=85mm]{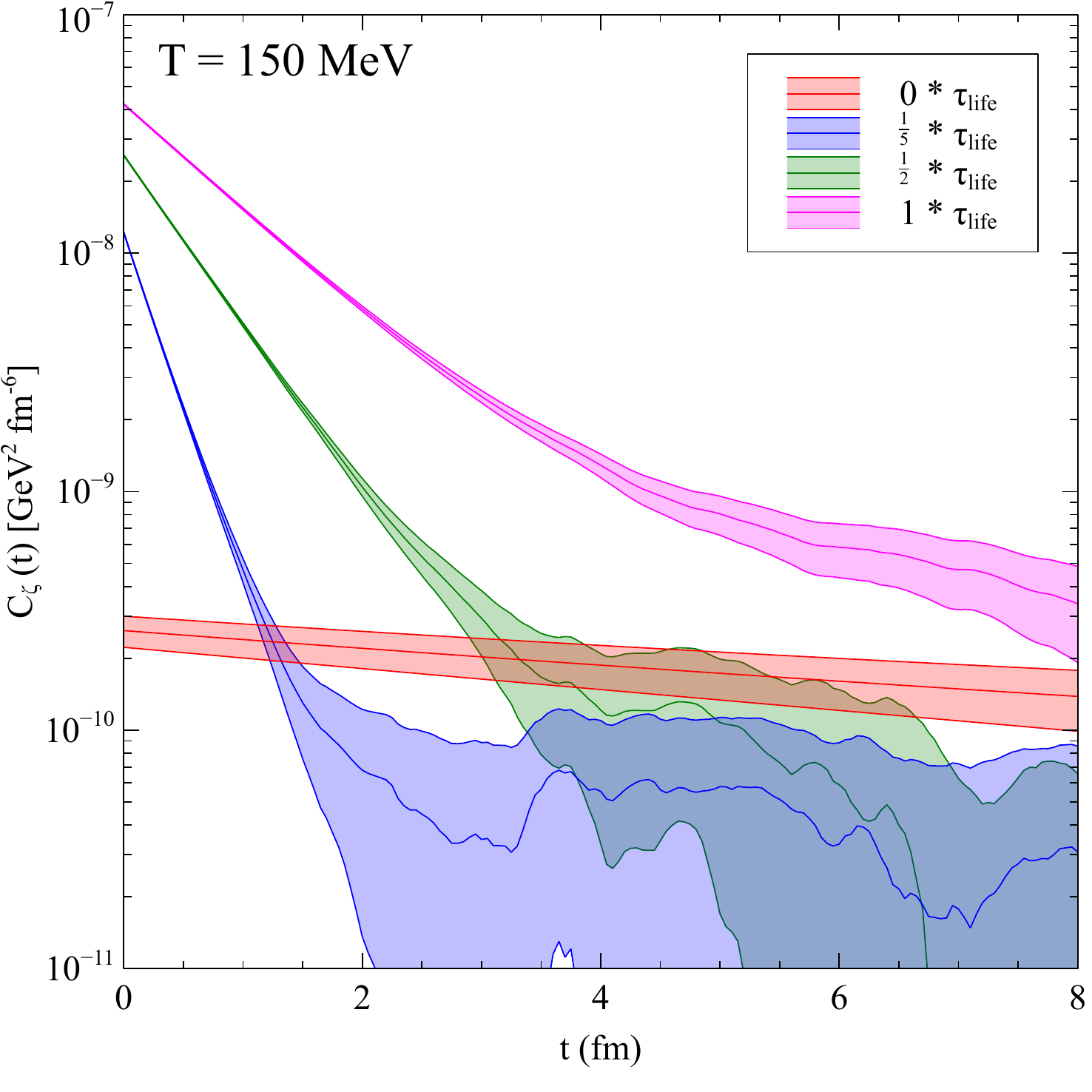}
  \caption{Bulk correlation function at $T=150$ MeV for a pion gas using the $\rho$ resonance for the cross-section for the cases where the $\rho$ has zero, a fifth, half or its full lifetime.}
  \label{fig:bulk_correls_pirho_lifetime}
\end{figure}

Figure~\ref{fig:bulk_correls_pirho_lifetime} shows the correlation function at $T=150$ MeV of the $\pi-\rho$ mixture in which the lifetime (and thus its relative abundance\footnote{In equilibrium, a fraction of the system is composed of transient $\rho$ resonances. If only the lifetime is modified without affecting the width used for the cross-section, then the scattering rate remains the same but the proportion of $\rho$ resonances will vary according to the same scaling as was applied to the lifetime.}) is varied and eventually taken to zero. In this precise limit we recover the previously discussed case of $2\leftrightarrow 2$ elastic scattering with no intermediate resonance. First notice that the correlation decays exponentially, as we previously assumed. Second, it is also evident that even a very small $\tau_{{\rm life}}$ profoundly modifies the underlying physics. Such a lifetime allows for a continuous formation and decay of a resonance allowing the imbalance in the longitudinal (bulk) channel to relax in a more effective way than a pure local collision. Such a mechanism produces a large increase of the fluctuations (seen in $C_\zeta (0)$), and a reduction of the relaxation time. By decreasing the resonance lifetime, we increase the number of decays and recombinations per unit time ($\pi+\pi \rightarrow \rho \rightarrow \pi+\pi$). Therefore the relaxation time of the bulk viscosity is shortened, as  can be seen in the figure. However, in the zero lifetime limit the collisions become effectively elastic, and we reach a limit in which the momentum relaxation is ineffective, with a very large relaxation time. 

This example shows the large dependence of $\tau_\zeta$ on the resonance lifetime. However, no deviation from the exponential form can be inferred so far.

\subsection{Several resonances: effect on correlation function}

Our previous analysis concerning the relaxation time dependence on the resonance lifetime was still possible on the basis of the single exponential decay of $C_\zeta(t)$. For such a system with a single channel (one resonance) the correlation function does not develop a non-exponential behavior similar to what we see in Fig.~\ref{fig:bulk_correls_full}. As the next step, it is possible to speculate that the presence of several interactions and decay modes, with a variety of relaxation times, determines the more complicated form of $C_\zeta(t)$. This idea was explored for a binary system of quarks and gluons in the BAMPS model in Ref.~\cite{El:2012ka} where a single exponential ansatz was unable to describe the shear viscosity correlation function.

To validate this hypothesis we study the effect of two independent resonant channels in the box. We start with the $\pi-\rho$ system of the previous section (using the physical $\rho$ lifetime, not modified anymore). In addition, we introduce a parallel particle/resonance system in the simulation. We add a non-physical particle species $B$ with the same mass as the pion ($m=138$ MeV) interacting through a single resonance $B^*$ with the same pole mass as the $\rho$ meson ($m=776$ MeV); however, we use a much smaller decay width for the $B^*$ (see Table~\ref{tab:particle_properties}). In summary, we insert a copy of the $\pi-\rho$ system but with a longer-lived resonance\footnote{For practical purposes the $B-B^*$ system is based on the kaon-$a_0$ system in SMASH with modified parameters according to Table~\ref{tab:particle_properties}. This notably means even though the masses and widths are identical, the degeneracies and recombination patterns of the particles and resonances are not exactly the same and can have an impact on $C_\zeta (0)$ and $\tau_\zeta$; this has been verified to be a small effect.}. Finally, to simplify the analysis, note that the $\pi-\rho$ and the $B-B^*$ are not coupled to each other.

\begin{table}[ht]
\begin{center}
 \begin{tabular}{||c | c | c| c||} 
 \hline
 Particle & Mass (MeV) & Width (MeV) & Decay channel \\ [0.5ex] 
 \hline\hline
 $\pi $ & 138 & 0 & -\\ 
 $\rho $ & 776 & 149 & $\pi +\pi$ \\ 
 $B $ & 138 & 0 & - \\ 
 $B^*$ & 776 & 20 & $B+B$ \\[1ex] 
 \hline
\end{tabular}
\caption{Properties of the species present in the $\pi-\rho-B-B^*$ system. The 2 resonances, $\rho$ and $B^*$ decay exclusively to 2 pions and 2$B$, respectively, with the total decay widths shown in the third column.}
\label{tab:particle_properties}
\end{center}
\end{table}

The reason for such a particular system is the following. Under a fluctuation in the bulk channel, the $\pi-\rho$ subsystem will have a relaxation time of the order of $\tau_{\rm life} \sim 1/\Gamma\simeq 1$ fm (similar to the result in the previous section). The new $B-B^*$ system has a significantly lower cross-section, and a lifetime which is an order of magnitude larger; this new subsystem is thus expected to relax $\sim 10$ times slower than the $\pi-\rho$ one, and it is expected that this separation of time scales will be visible in the correlation function of the mixture.

\begin{figure}[ht]
  \centering
  \includegraphics[width=76mm]{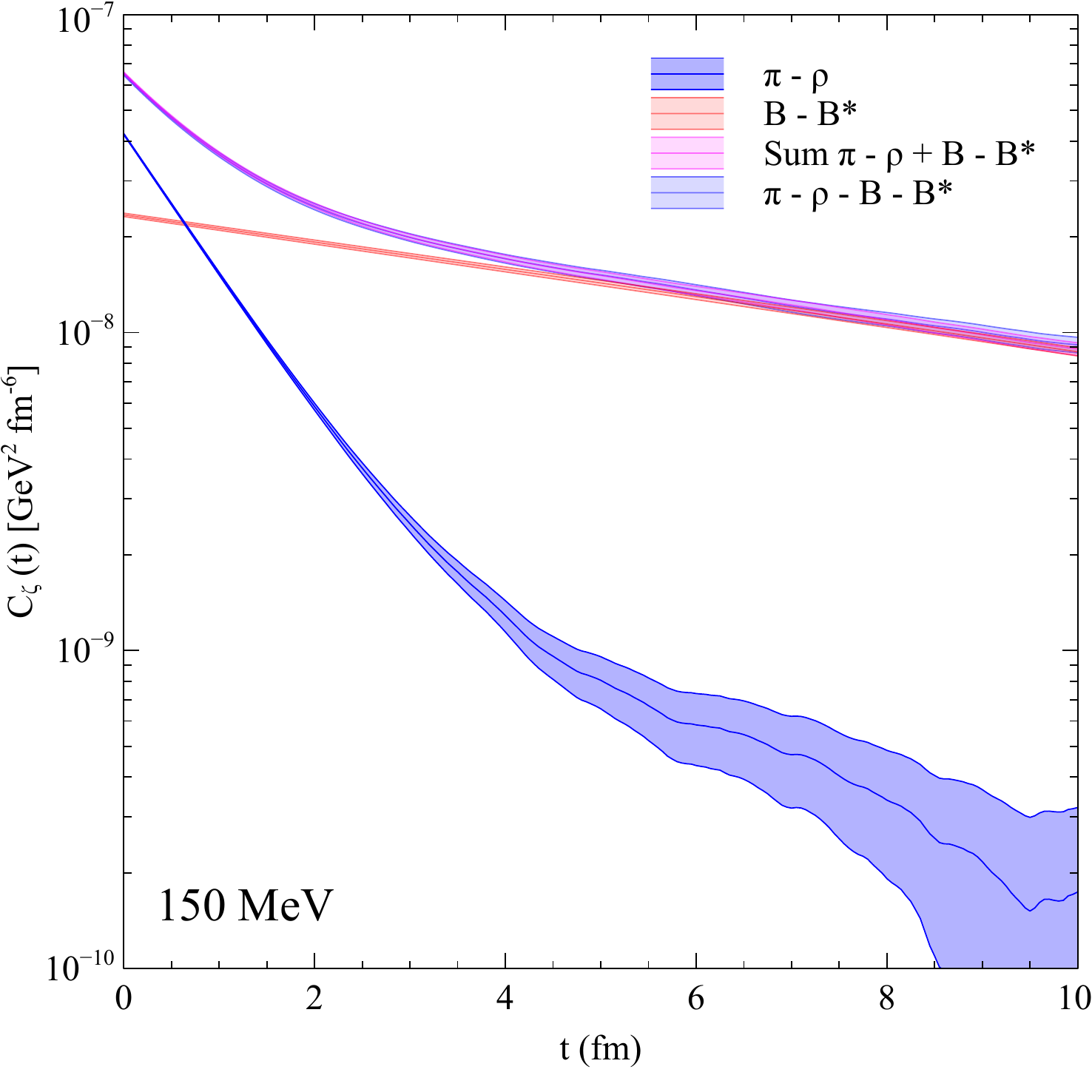}
  \includegraphics[width=84mm]{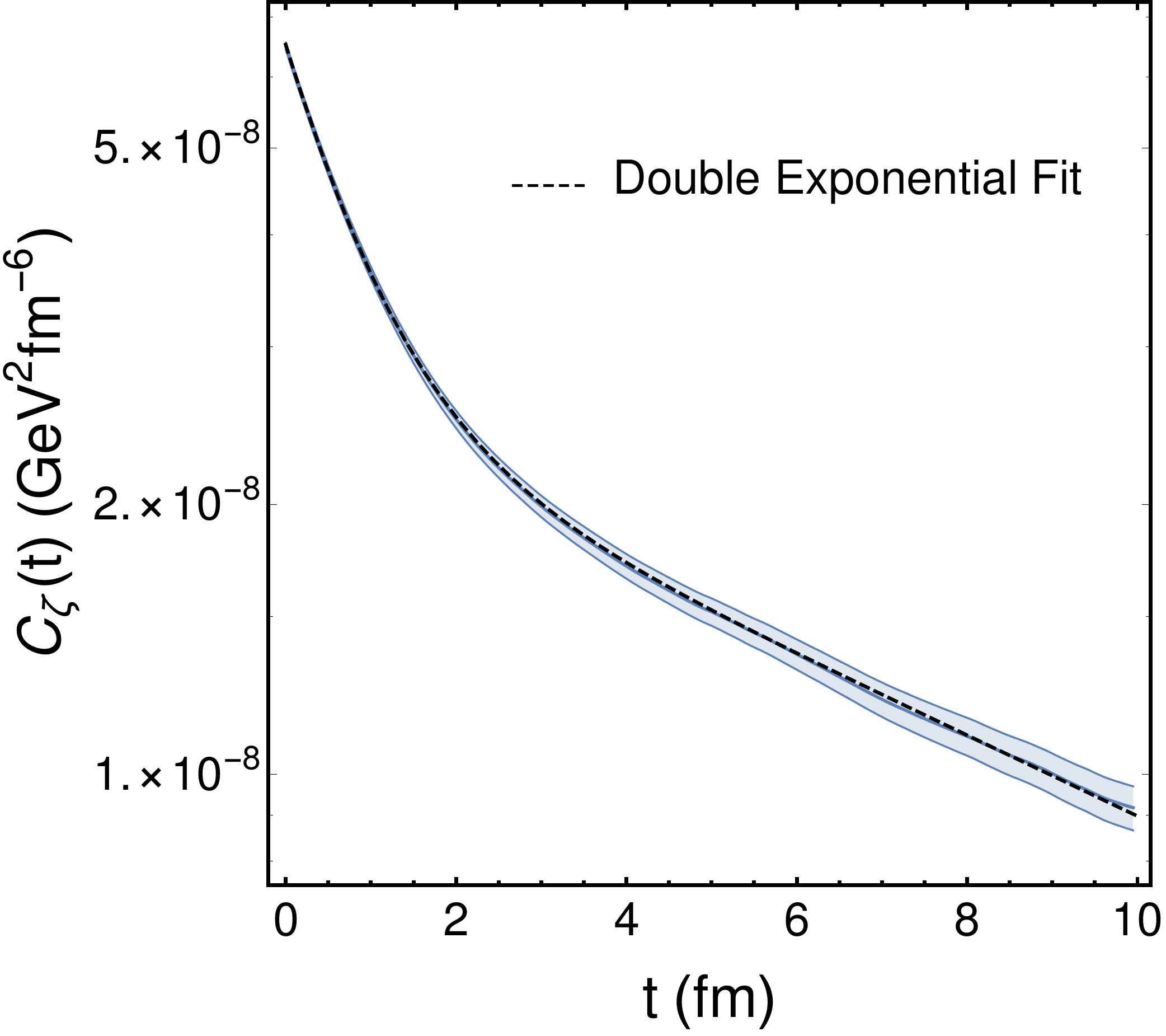}
  \caption{Left: Bulk correlation functions for various gases interacting through resonances, $\pi-\rho$, $B-B^*$, and $\pi-\rho-B-B^*$ at $T=150$ MeV. The direct sum of correlation functions of $\pi-\rho$ plus the $B-B^*$ is indistinguishable from the correlation function of the mixture $\pi-\rho-B-B^*$. Right: Double exponential fit of the mixture of gases.
  }
  \label{fig:bulk_correls_pirhoKa0}
\end{figure}

We plot the bulk correlation function of the different systems in the left panel of Fig.~\ref{fig:bulk_correls_pirhoKa0}. As expected, the $\pi-\rho$ subsystem (in blue) has a smaller relaxation time than the $B-B^*$ system (flatter red curve). The smaller $C_\zeta(0)$ of the $B-B^*$ is due to the more suppressed resonant contribution, as the broader $\rho$ resonance weights more in the thermodynamic average of $C_\zeta (0)$. For both subsystems the correlation function is a single exponential, as expected.

Looking at the correlation function of the mixture of $\pi-\rho-B-B^*$, we observe a non-exponential shape comparable to the ones for the full hadron gas in Fig.~\ref{fig:bulk_correls_full}. Even more interestingly, adding up the individual exponential contributions from the $\pi-\rho$ and $B-B^*$ systems results very precisely in the same correlation function for the full system, with later times being dominated by the $B-B^*$ process, the slowest one.

We proceed to fit the resulting correlation function of the mixture at $T=150$ MeV to a double exponential form,
\be C_\zeta(t) = C_{\zeta,\pi} (0) e^{-t/\tau_{\zeta,\pi}}+C_{\zeta,B} (0) e^{-t/\tau_{\zeta,B}} \ . \ee
The tail of the correlation function is first fitted to extract $C_{\zeta,B}(0)$ and $\tau_{\zeta,B}$, and then subtracted from the total correlation function. After checking that the remaining function is indeed exponential, it is fitted to obtain $C_{\zeta,\pi}(0)$ and $\tau_{\zeta,\pi}$. The final fit is shown in the right panel of Fig.~\ref{fig:bulk_correls_pirhoKa0} in dashed line (the correlation function itself is hidden by the fit, but its error band is still visible). We obtain $\tau_{\zeta,\pi}=0.91$ fm and $\tau_{\zeta,B}=9.65$ fm. These values turn out to be of the same order of magnitude as the respective lifetimes $1/\Gamma_\rho \simeq 1.32$ fm and $1/\Gamma_{B^*} \simeq 9.85$ fm. Thus, both microscopic processes of resonance formation/decay do affect the bulk viscosity of the mixture, each of them with its own characteristic relaxation time.

It is therefore natural to expect that the full hadron gas, being a massively more complex system, will be described by a collection of individual exponentials. However, contrarily to the case we just discussed, since many of the subsystems of the full hadron gas are actually coupled to each other, it would be complicated to associate these individual exponentials to a specific individual particle-resonance pair. Each one will correspond to each of the many interlinked subsystems (containing elastic and/or inelastic processes) present in the gas, with later times being dominated by the slowest such subsystem (i.e. the one containing the slowest set of processes).

In a more general way, one should then replace the single exponential ansatz by a linear combination of many such exponentials,
\be C_\zeta(t)= \int_0^\infty d\tau \rho(\tau) \exp \left( -t/\tau \right) \ , \label{eq:Ct} \ee
where the kernel function of relaxation times $\rho(\tau)$ is normalized to $\int_0^\infty d\tau \rho(\tau)=C_\zeta(0)$ and can be found, in principle, via a Laplace transform of the correlation function~\cite{gardner1959method}. Notice that the range of the possible relaxation times runs from 0 to $+\infty$, accommodating fast as well as slow processes.

However, in the remaining part of this work we do not need to use the full integral version of Eq.~(\ref{eq:Ct}), as we will see that the kernel function $\rho(\tau)$ can be taken as a linear combination of a few Dirac deltas,
\be \label{eq:shah} \rho(\tau) = 2 \sum_i^N C_{\zeta,i} (0) \delta(\tau-\tau_{\zeta,i}) \quad , \quad \sum_i^N C_{\zeta,i} (0)=C_{\zeta} (0) \ , \ee
one for each relaxation time taking place in the system. Notice that for $N=1$ one recovers Eq.~(\ref{eq:correl_ansatz}).

\section{Full hadron gas~\label{sec:hadrongas}}

We focus again on the correlation functions of Fig.~\ref{fig:bulk_correls_full} for the full hadron gas, and use the multi-exponential form \eqref{eq:Ct} and \eqref{eq:shah} to fit them. By inspection, we observe that $N=3$ components (that is, three Dirac deltas) are sufficient to achieve a good fit of the correlation functions. The corresponding relaxation times should be considered as the dominant modes contained in the kernel $\rho(\tau)$, which are related to physical processes in the hadron system. Of course, many other relaxation times do exist in the mixture (in principle, as many as independent microscopic processes), but they carry such a small amplitude that are not reflected in the correlation function. We perform a global fit using the ROOT library, which takes into account the error band of $C_\zeta(t)$ and provides statistical uncertainties of the fitting parameters. A much more detailed discussion on the multi-exponential fitting can be found in App.~\ref{app:fitting}.

We provide the results for two different sizes of the box, denoted as ``larger'' and ``smaller box'', the precise lengths of which are given in Table~\ref{tab:box}. 

\begin{table}[ht]
\begin{center}
\begin{tabular}{|c||c|c|} 
 \hline
 $T$ (MeV) & larger box, length & smaller box, length \\ 
\hline
\hline
86 &100 fm & 60 fm \\
114 &  60 fm & 40 fm \\
142  &  40 fm & 20 fm \\
172  &  20 fm  & 10 fm \\
\hline
\end{tabular}
\caption{Box sizes for the full hadron gas in SMASH. For each temperature we use two sets of volumes, larger and smaller. The volume of each box is $V={\rm length}^3$.
}
\label{tab:box}
\end{center}
\end{table}

\begin{figure}[ht]
  \centering
 \includegraphics[width=80mm]{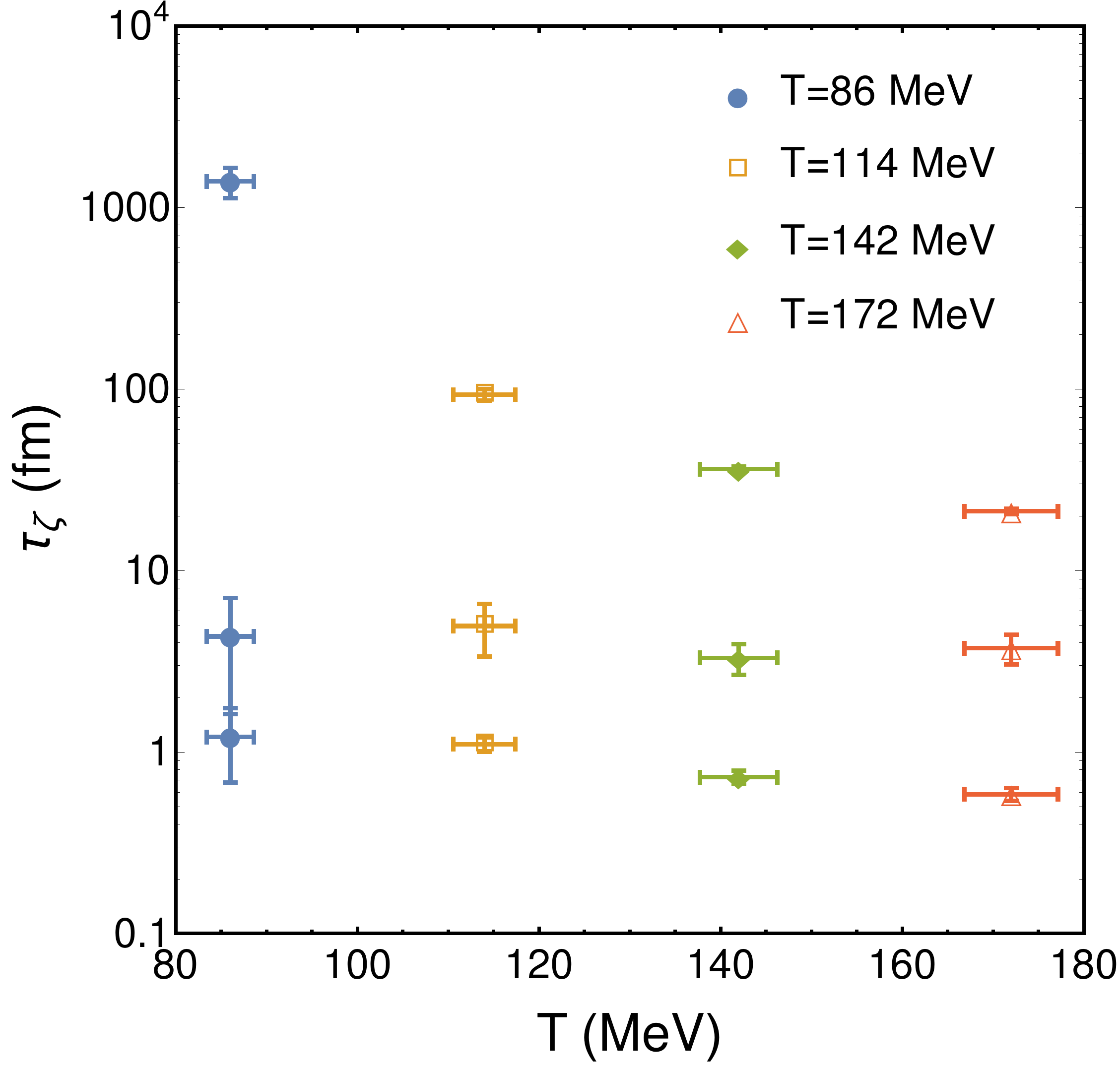}
    \caption{Relaxation times from 3-mode fits for the full hadron gas in SMASH at four different temperatures in the larger box calculation.}
  \label{fig:relax_times}
\end{figure}

The resulting fitting parameters are summarized in Tables~\ref{tab:fits} and \ref{tab:fits2} of App.~\ref{app:fitting}. Among them, the relaxation times for each temperature in the larger box are given in Fig.~\ref{fig:relax_times}, where one can see the different time scales taken place in the gas dynamics.

%

\begin{figure}[ht]
  \centering
 \includegraphics[width=80mm]{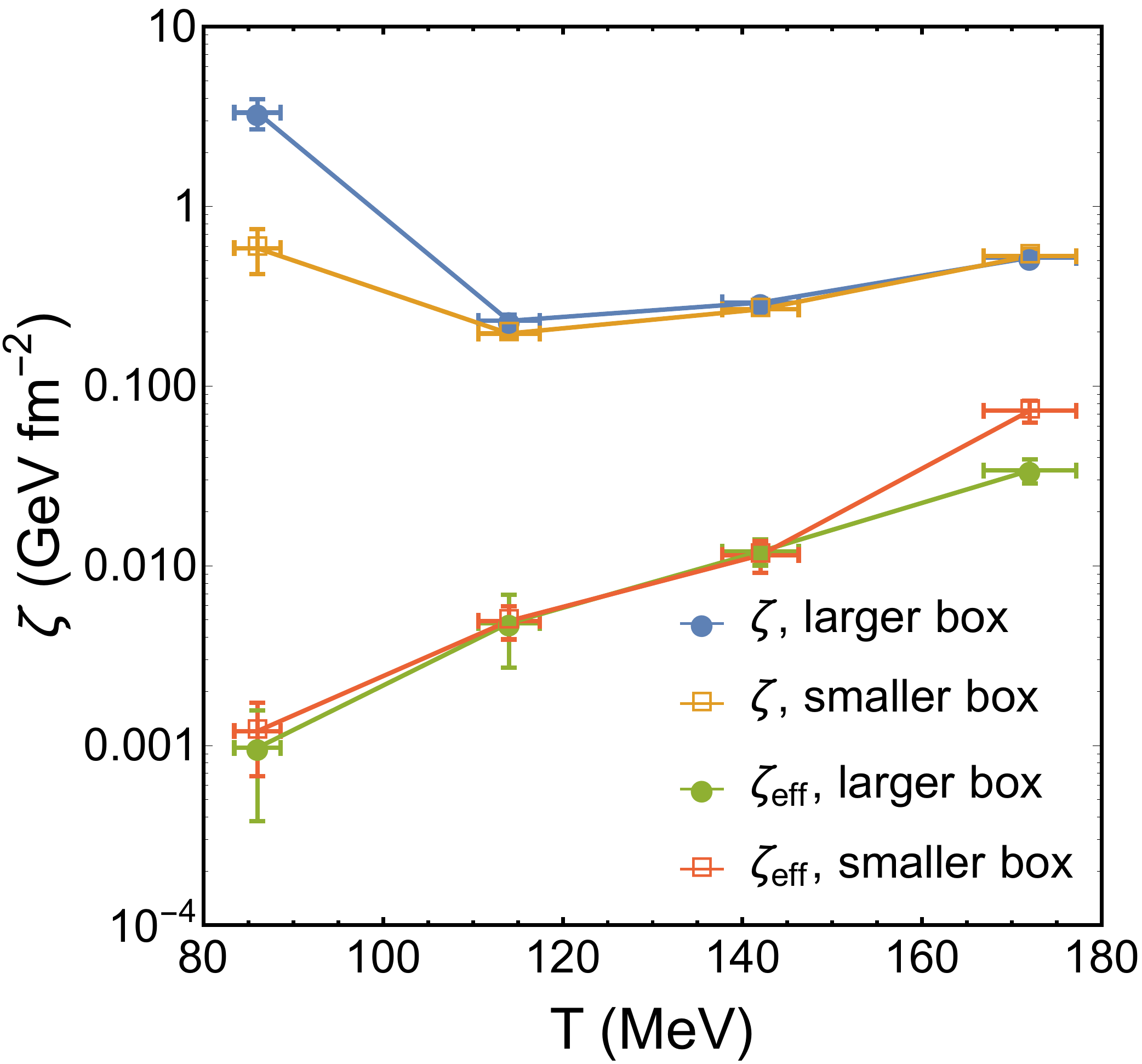}
 \includegraphics[width=80mm]{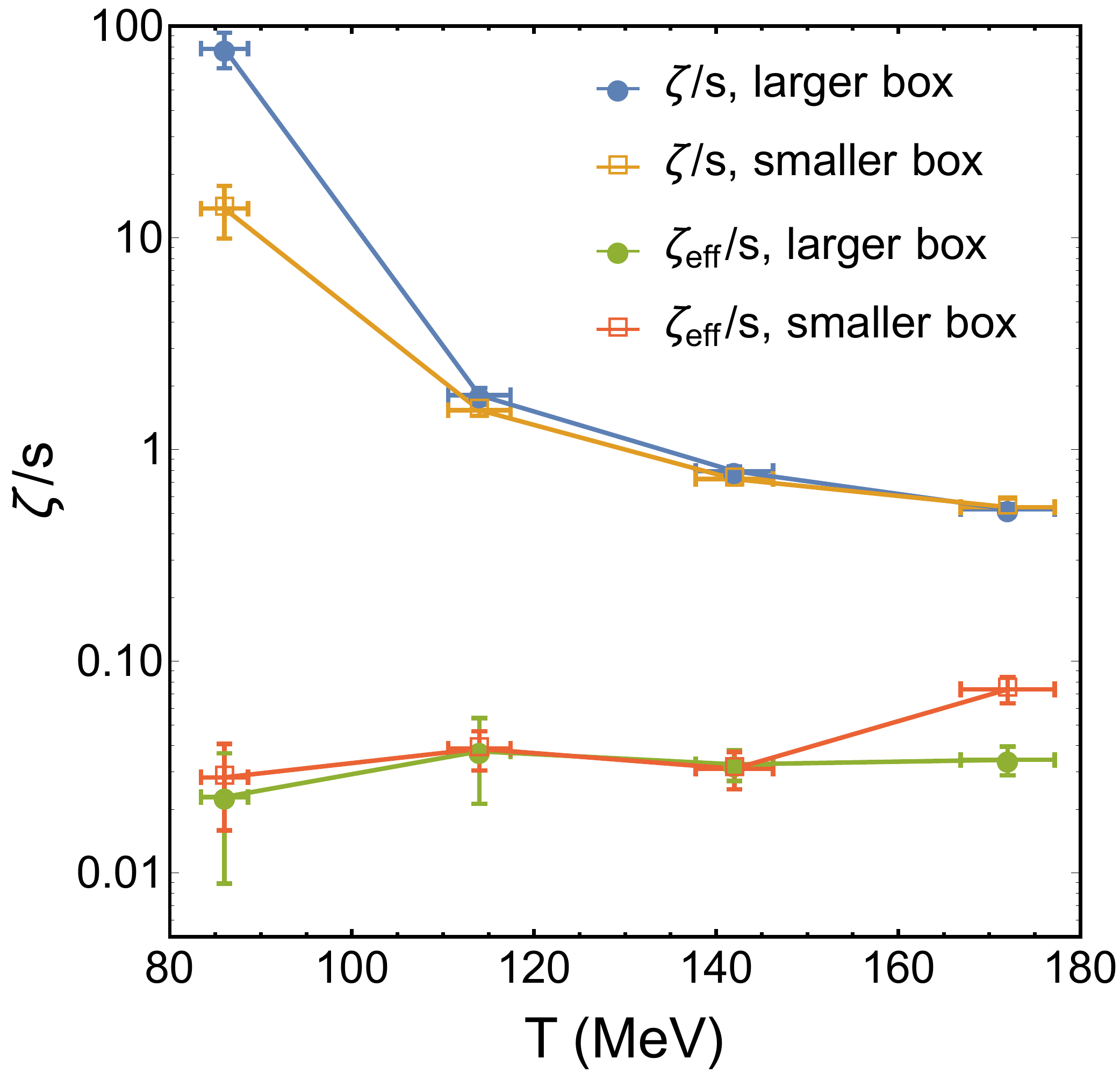}
  \caption{Bulk viscosity for the full hadron gas for 2 different box sizes. As explained in the text, the effective bulk viscosity is obtained by removing the long-lived modes.
  Left panel: Bulk viscosity $\zeta$. Right panel: $\zeta/s$. The different box sizes are described in Table~\ref{tab:box}.
  }
  \label{fig:bulk_correls_schemes}
\end{figure}

In Fig.~\ref{fig:bulk_correls_schemes} we plot the final result of the bulk viscosity (left panel) and the bulk viscosity over entropy density (right panel) for the full hadron gas as functions of the temperature.

Along with this ``inclusive'' $\zeta$ and $\zeta/s$ (where all modes present in the fit are included in the calculation) we have included results of an ``effective'' bulk viscosity coefficient. The latter is calculated by taking the long-lived modes out of the analysis for phenomenological reasons: In an infinite-lived system all modes contribute to the correlation function at some point, as the total relaxation of a fluctuation does not happen entirely until all modes in the system have equilibrated. In particular, the slowest mode is typically the one that dominates the bulk viscosity, as it is the one describing the long tail of the correlation function; however such slow processes are not effective in a short-lived system, if their inverse rate is much larger than the lifetime of the system. If in RHICs the hadronic phase lasts approximately 10-30 fm/$c$, then a relaxation mode with $\tau_\zeta=10^2-10^3$ fm cannot play any role. The part of the system corresponding to that mode remains out of equilibrium for the whole time, and does not contribute to the transport coefficient calculation. We define the effective bulk viscosity $\zeta_{\rm{eff}}$ to be the transport coefficient where such modes have been excluded.

More formally, the effective bulk viscosity can be defined as 
\be \label{eq:eff}
\zeta_{ \textrm{eff}} = \frac{V}{T} \int_{0}^{\infty} dt \ C_{\zeta,\textrm{eff}} (t,\tau^*) \ ,
\ee
where the effective correlation function now depends on a cutoff $\tau^*$, or the order of the lifetime of the system, above which the modes are suppressed. Using e.g. a hard cutoff to remove these modes,
\be C_{\zeta,\textrm{eff}}(t,\tau^*)= \int_0^{\tau^*} d\tau \rho(\tau) \exp \left( -t/\tau \right) \ . \ee
Note that to obtain the effective bulk viscosity one still integrates the correlation function  up to $\infty$, but the kernel $\rho(\tau)$ is restricted. This definition still assumes the validity of the exponential ansatz for every mode.  

Why should this effective bulk viscosity be of any relevance? Suppose that one tries to describe the evolution of the system by a relativistic hydrodynamic code for heavy-ion collisions with the bulk viscosity as an input parameter to be fixed~\footnote{In practice the late dilute stage of the hadronic evolution is usually simulated via a transport approach, instead of using hydrodynamics, but the argument is equally valid for the denser part of the evolution described by hydrodynamics.}. We argue that the extremely long-lived processes will hardly happen during the real evolution of the system, so they cannot be part of the eventually-inferred viscosity. The effective transport coefficient defined here should be associated to the one obtained from matching experimental observables using hydrodynamic codes; in contrast, the inclusive bulk viscosity should rather be compared with a theoretical calculation, e.g. solving the Boltzmann equation in the thermodynamic limit. Due to the suppression of the dominant mode (or modes), it is clear that $\zeta_{{\rm eff}}$ should always be smaller than $\zeta$.

To completely understand this distinction, let us finally present another example of such an effective viscosity. Mannarelli et al.~\cite{Mannarelli:2012eg} calculated the shear viscosity due to phonons in optically trapped cold Fermi atoms. At low temperatures, the mean free path of phonons increases and exceeds the physical boundaries of the superfluid region. The shear viscosity is proportional to the mean free path, so at low temperatures it is possible to define an effective shear viscosity where the mean free path is replaced by a distance of the order of the atomic cloud. In our particular case, the bulk viscosity is proportional to a linear combination of relaxation times, and the effective bulk viscosity imposes a limiting time of the order of the system's own duration.

In our study, motivated by RHIC physics, we calculate this effective bulk viscosity by removing the slowest mode of the three (upper points in Fig.~\ref{fig:relax_times}), whose relaxation time is typically much larger than the hadronic lifetime in a RHIC. Notice that for the highest temperatures $T = 142$ MeV and $T=172$ MeV, $\tau_{\zeta,3}$ is actually of the order of the lifetime of the fireball, and one could argue that this mode can still play some role in heavy-ions. Therefore, one should strictly interpret the effective bulk viscosity as a lower bound in these cases. Also note that this implies that systems with different lifetimes could have a different effective bulk viscosity, such as for example in the experiments at the very different beam energies of the Relativistic Heavy Ion Collider and the Large Hadron Collider.

The final results for $\zeta$ and $\zeta/s$ in Fig.~\ref{fig:bulk_correls_schemes} behave similarly for both box volumes. $\zeta/s$ decreases systematically with temperature to values around $\zeta/s \simeq 1$, perhaps reaching a plateau around $T=172$ MeV. Only the result for $T=86$ MeV is quite different in the two volumes, and might correspond to a poor quality in one of the volumes used, similar to the discrepancy in the simple pion gas in Sec.~\ref{sec:bulk_pion}. 

The effective bulk viscosity is always smaller in magnitude, as expected. The $\zeta_{{\rm eff}}$ at $T=172$ MeV is somewhat different between the two volumes, due to the different value of the $\tau_{\zeta,2}$ for that temperature. It is not evident to us which one of the two, if any, is of lesser quality.  $\zeta_{{\rm eff}}/s$ is a rather flat or slightly increasing function of the temperature for the considered range. 
\begin{figure}[ht]
  \centering
 \includegraphics[width=80mm]{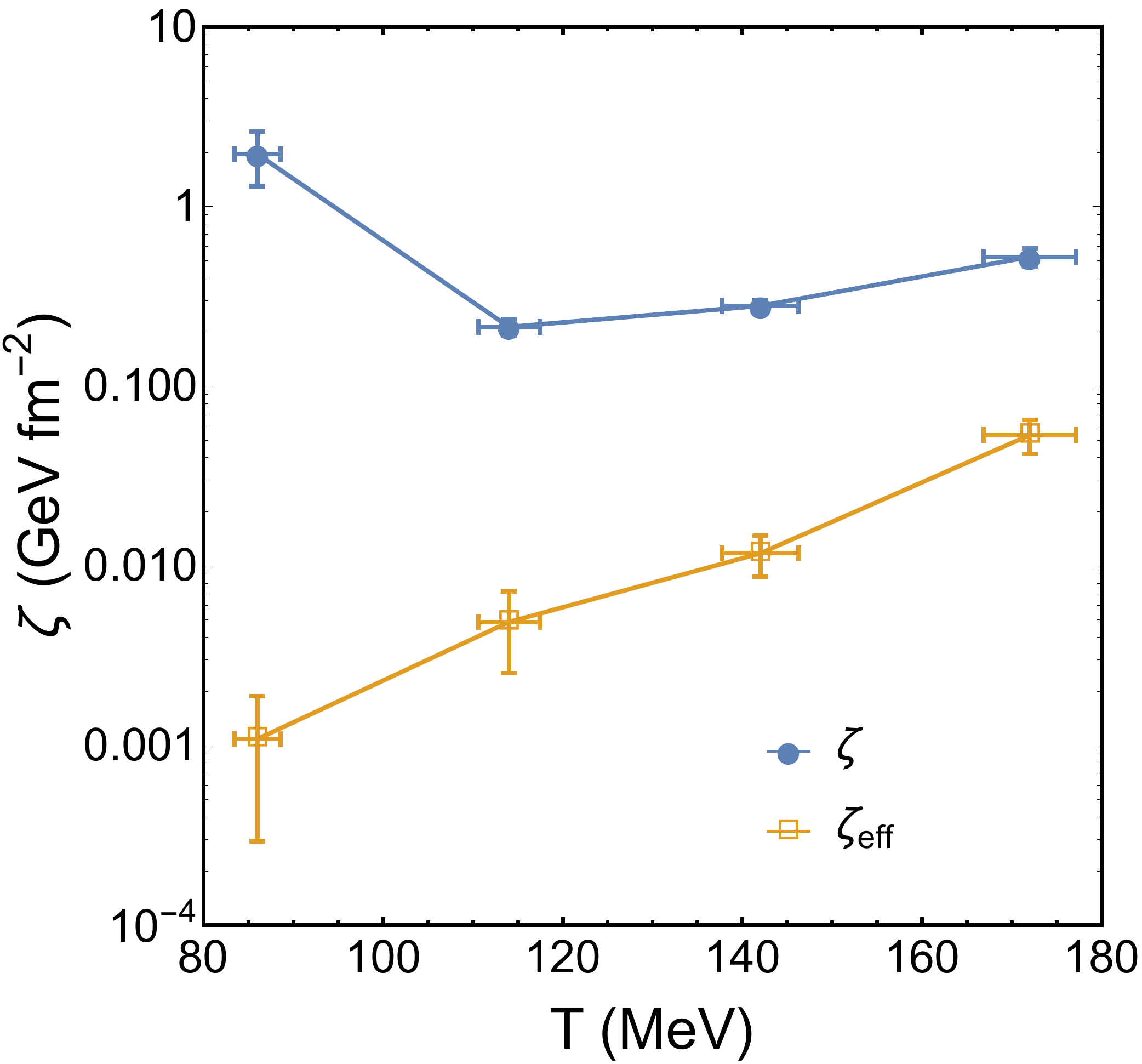}
 \includegraphics[width=80mm]{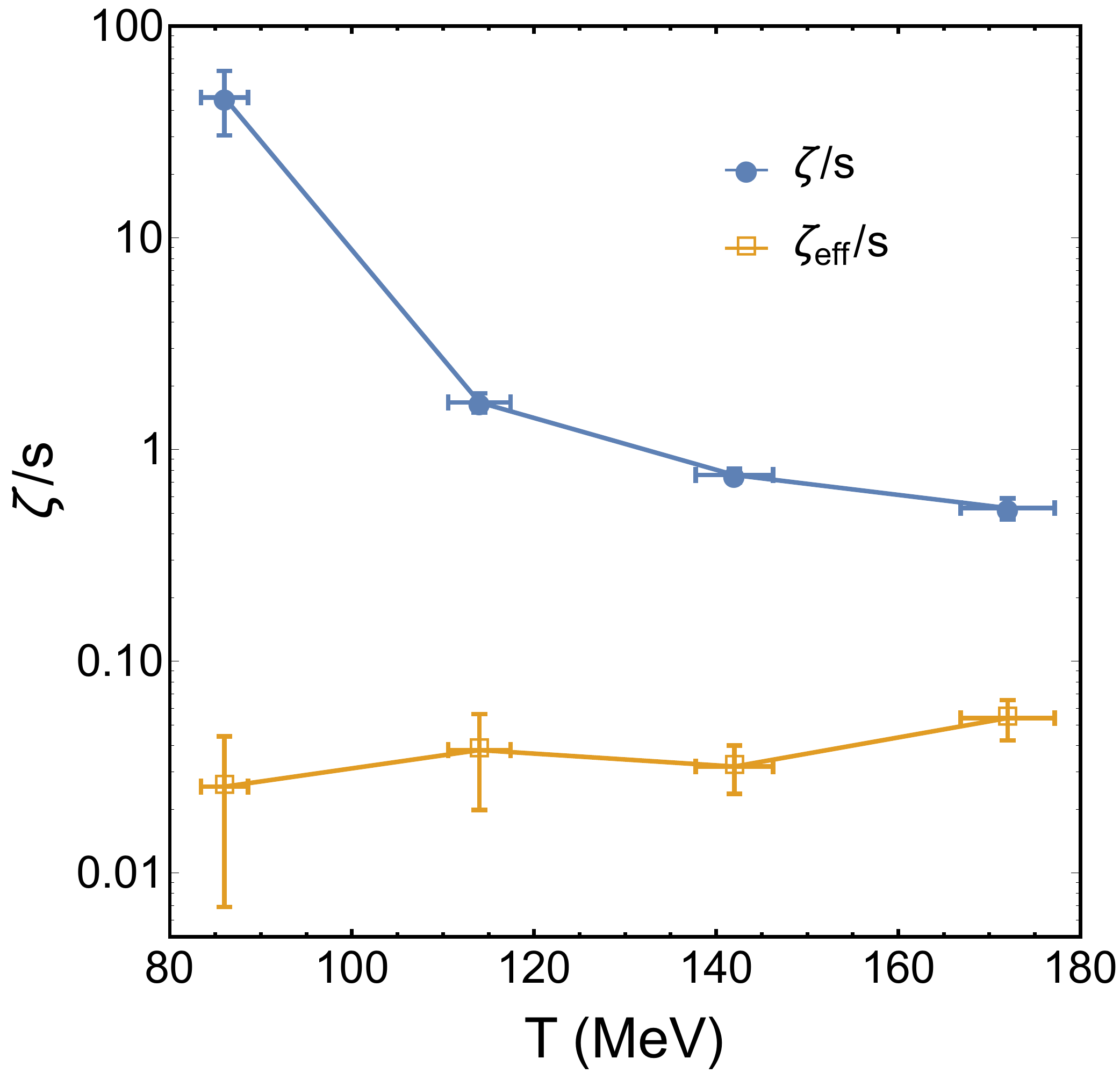}
  \caption{Averaged (over 2 box sizes) bulk viscosity for the full hadron gas. Left panel: $\zeta$ and $\zeta_{{\rm eff}}$. Right panel: $\zeta/s$ and $\zeta_{{\rm eff}}/s$.}
  \label{fig:bulkfinal}
\end{figure}

The final value of our coefficients is obtained by averaging the two box sizes and combining their uncertainties. Our average value for $\zeta (\zeta/s)$ and $\zeta_{{\rm eff}} (\zeta_{{\rm eff}}/s)$ is shown in the left (right) panel of Fig.~\ref{fig:bulkfinal}.

\subsection{Discussion and comparison}

In this section we attempt to contextualize the present calculation by testing it against previous calculations of the bulk viscosity. Before doing so, let us briefly comment on the relation between the bulk viscosity and the adiabatic speed of sound $v_S$ defined in Eq.~(\ref{eq:vs}).

In a massless, weakly-coupled gas, previous calculations using the Boltzmann equation and kinetic theory have shown that the relation between shear and bulk viscosity should be proportional to the squared non-conformality parameter~\cite{Weinberg:1971mx,Hosoya:1983id,Horsley:1985dz,Arnold:2006fz}
\begin{equation}
\frac{\zeta}{\eta} \simeq 15 \left( \frac13 - v^2_S \right)^2 \ .
\label{eq:conformality} 
\end{equation}

\begin{figure}[ht]
  \centering
  \includegraphics[width=70mm]{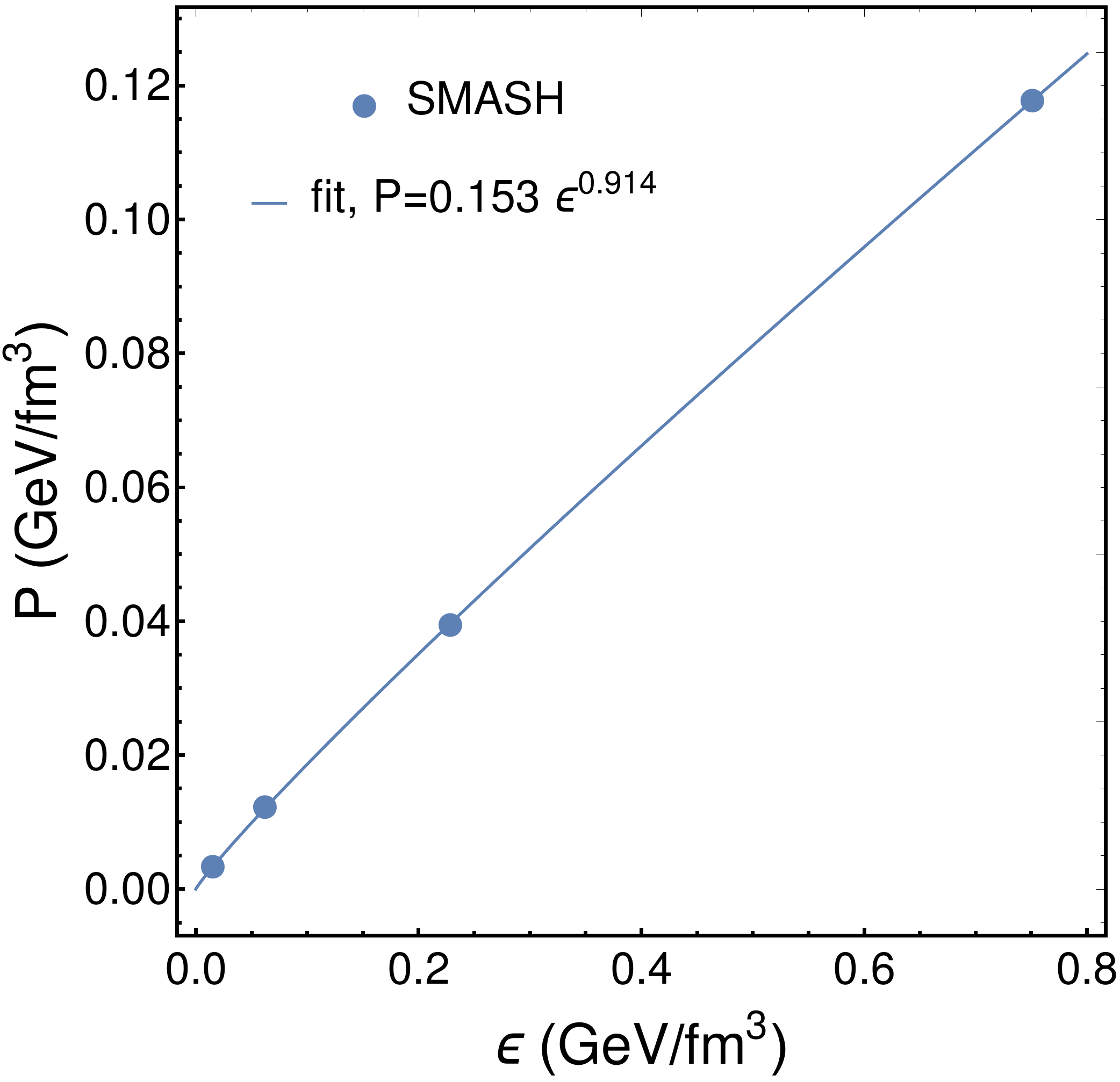}
   \includegraphics[width=70mm]{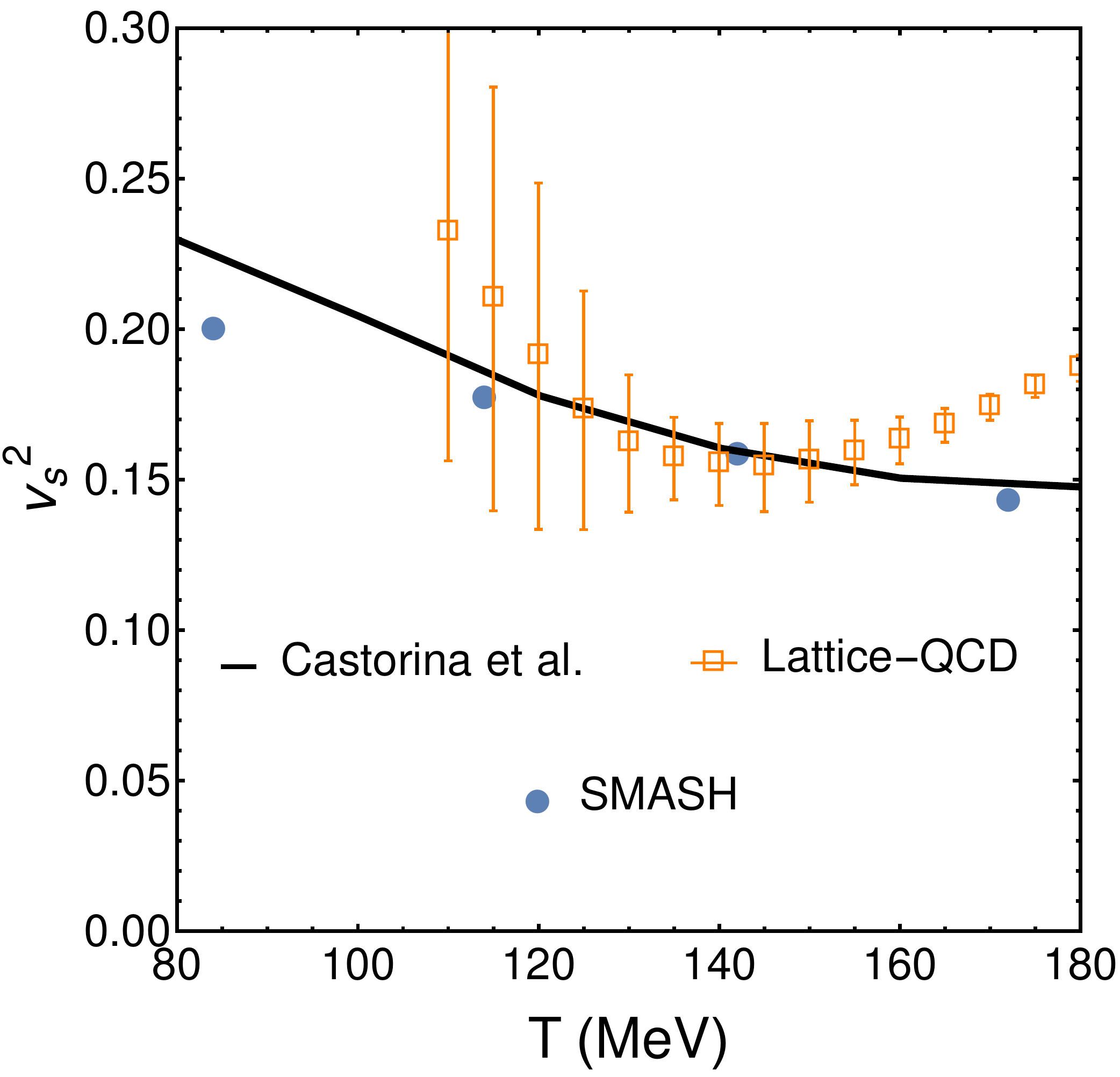}
\caption{Left panel: Pressure and energy density for each of the 4 temperatures in the full hadron gas computed in SMASH (line is a fitting function in the form $P=a\epsilon^b$) . Right panel: Speed of sound versus temperature. SMASH result are obtained by using the fit made in the left panel. The line is the result of the physical resonance gas approximation in~\cite{Castorina:2009de}, and the lattice-QCD data is extracted from Ref.~\cite{Borsanyi:2013bia}.}
  \label{fig:speedsound}
\end{figure}

\begin{figure}[ht]
  \centering
   \includegraphics[width=80mm]{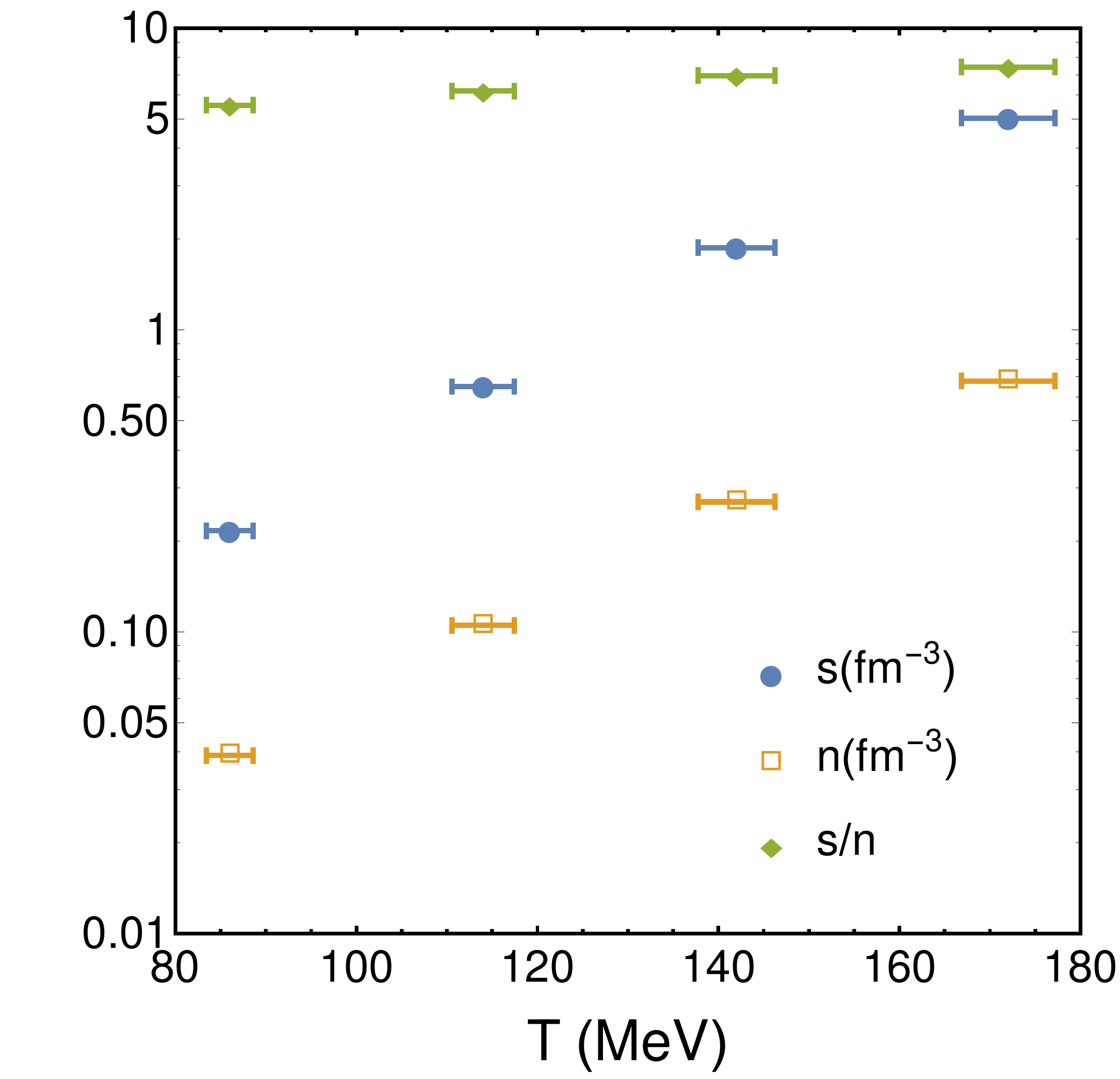}
\caption{Entropy density, number density, and entropy per particle of each of the four temperature points, simulated in the full hadron gas using SMASH.
}
  \label{fig:sovern}
\end{figure}

We can try to estimate the adiabatic speed of sound from the measurements of the energy density and pressure in SMASH for the full hadron gas studied before. This is illustrated in the left panel of Fig.~\ref{fig:speedsound}, where we plot the values of these two quantities for each of the four temperatures. This shows the dependence of $P$ versus $\epsilon$ needed to obtain the speed of sound. Before extracting $v_S^2$, we verify whether the entropy density $s$, number density $n$ or entropy per particle $s/n$ is held constant in these measurements, as we have not imposed any of those conditions explicitly. This is detailed in Fig~\ref{fig:sovern}, where we provide the values of $s$, $n$ and $s/n$ for each temperature. None of the three quantities remains absolutely constant but it is clear that one can rule out an isentropic (constant $s$) and isochoric (constant $n$) dependence. On the other hand, the entropy per particle does not vary much. Therefore it is fair to assume that the speed of sound, obtained from the relation between $P$ and $\epsilon$ in our plot, will be approximately adiabatic (constant $s/n$), or, at least, a close proxy for it.

%

We parametrize the dependence of the pressure to the energy density with a power law, and find that $P(\epsilon)=0.153 \ \epsilon^{0.914}$, where both quantities are measured in GeV/fm$^3$. The fit is shown as a solid line in the left panel of Fig.~\ref{fig:speedsound}. In the right panel of the same figure we show the resulting $v_S^2$ from this relation in blue dots, which is a decreasing function within this range of temperatures. Our values compare well with the result of Ref.~\cite{Castorina:2009de} for a hadron resonance gas including resonances up to a mass of 2.5 GeV (similar to ours), and it is also comparable with the lattice QCD calculation of Ref.~\cite{Borsanyi:2013bia}, the deviation at high temperatures being due to the absence of a deconfined phase in our model.

\begin{figure}[ht]
  \centering
  \includegraphics[width=110mm]{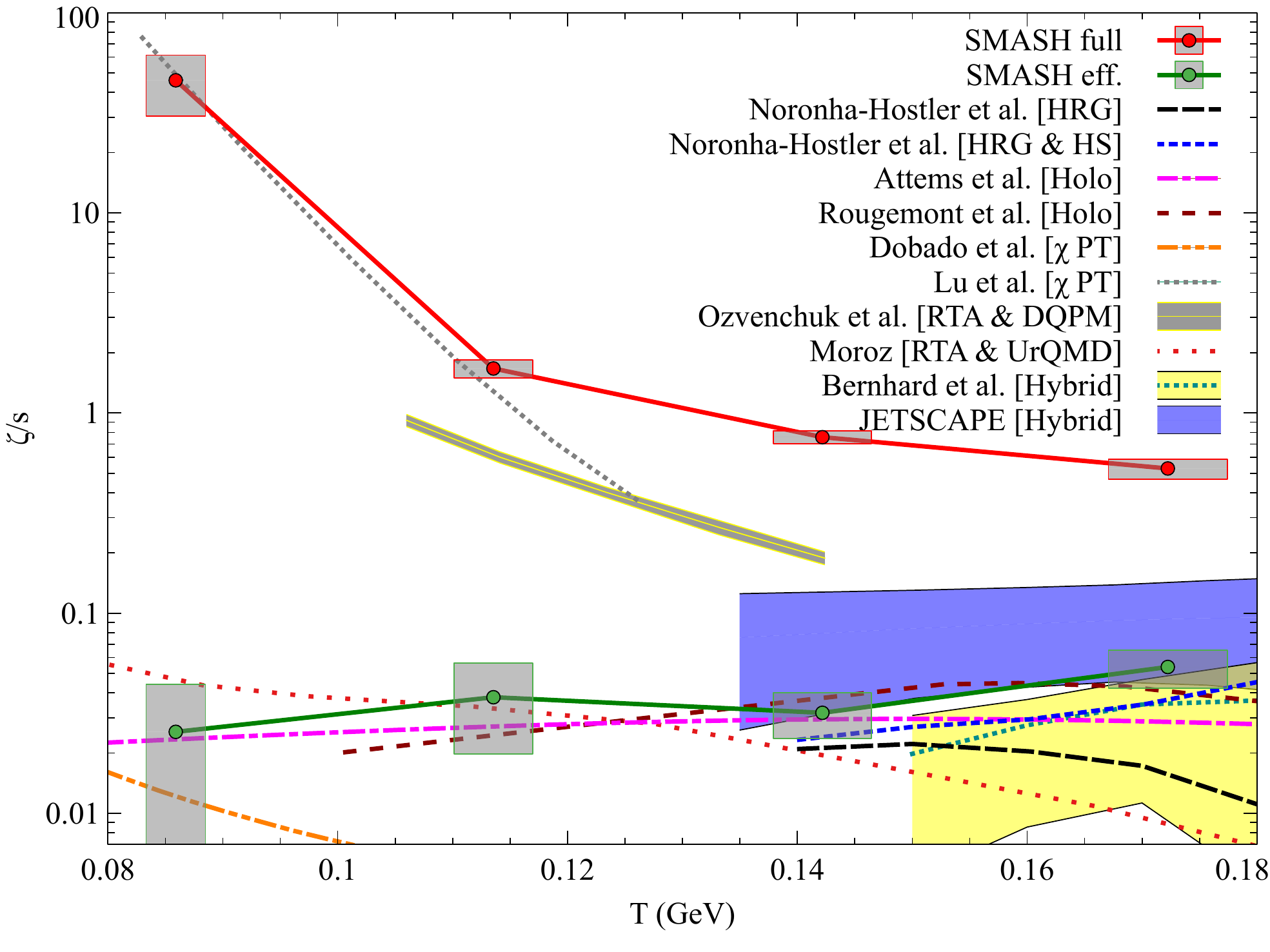}
  \caption{Comparison of several calculations for $\zeta/s$ at $\mu_B = 0$; see text for details.}
  \label{bulk_comparison}
\end{figure}

Finally we move to a comparison of available calculations for $\zeta/s$ presented in Fig.~\ref{bulk_comparison}, and we shortly discuss every other result with our own.
We have shown both $\zeta/s$ and $\zeta_{{\rm eff}}/s$ computed from SMASH in red and green symbols, respectively.

\begin{itemize}
 \item Noronha-Hostler et al.~\cite{NoronhaHostler:2008ju} use a hadron resonance gas model which assumes a comparable set of hadronic states as the ones used by SMASH. However, this model assumes a noninteracting tower of states, and the hadron resonance gas is supplemented with an exponentially increasing density of Hagedorn states [HS]. The bulk viscosity is calculated using the small-frequency spectral ansatz presented in~\cite{Karsch:2007jc}, which matches the Euclidean version of the correlator of the trace of the energy-momentum tensor. Their result is comparable to our $\zeta_{{\rm eff}}/s$, as that calculation lacks of the very slow dynamical process affecting our viscosity. An increase of $\zeta/s$ close to $T_c$ is only obtained by the inclusion of the Hagedorn states (see the two curves on Fig.~\ref{bulk_comparison}), and such an increase is not captured by any other model, except perhaps, by the SMASH effective bulk viscosity due to the amount of different resonances implemented here. Nevertheless, this increase turns out to be very mild, unless an actual phase transition with a rapid change in degrees of freedom is considered in the model close to $T_c$.

 \item Both Attems et al.\footnote{Note that the data in \cite{Attems:2016ugt} was obtained using a 150 MeV crossover temperature and $\phi_M =20$.}~\cite{Attems:2016ugt} and Rougemont et al.~\cite{Rougemont:2017tlu} calculations are performed in holographic setups, with certain degree of conformality breaking to get a finite bulk viscosity. Given the radical different framework it is difficult to compare to our own results. The bulk viscosity is small in these calculations and rather flat (with a broad peak close to $T=150$ MeV) which is compatible with our values of $\zeta_{{\rm eff}}/s$.

 \item The calculation of Dobado et al.~\cite{Dobado:2011qu} and Lu et al.~\cite{Lu:2011df} are both computed for a pure pion gas using chiral perturbation theory at low temperatures. However, their different approaches illustrate the conceptual difference between $\zeta$ and $\zeta_{{\rm eff}}$. While~\cite{Lu:2011df} considers the slowest number changing process affecting the bulk viscosity ($2 \pi \leftrightarrow 4\pi$) and neglects any elastic collisions, the calculation in~\cite{Dobado:2011qu} does not consider this process and uses the $2\pi \leftrightarrow 2\pi$ process only with a pion pseudochemical potential. In the first calculation the extremely slow inelastic process (suppressed by the derivative coupling at low energies) describes $\zeta$. In the second calculation these processes are absent during the hadronic stage of RHICs and only elastic collisions are able to build a $\zeta$ at the expense of the change in chemical potential. This might explain why~\cite{Dobado:2011qu} is closer to $\zeta_{\rm eff}/s$, while~\cite{Lu:2011df} is closer to $\zeta/s$. 
  
However one should also note that neither of these theoretical calculations include dynamical resonances like SMASH, and any agreement is probably accidental, as the scattering processes are different in the three calculations.

\item Interestingly, the PHSD calculation from Ozvenchuk et al.~\cite{Ozvenchuk:2012kh} is not far from our $\zeta/s$, which can be explained in part because PHSD also propagates resonances, and thus including mass changing processes. Using their discrete test particle representation, the bulk viscosity is computed from a discretized version of the relaxation time approximation,
\be
\zeta = \frac{1}{9TV} \sum_i \int \frac{d^3p}{(2\pi)^3} \frac{\Gamma_i^{-1}(p)}{E_{p,i}^2} \left( (1-v^2_S) E_{p,i}^2 - m^2_i \right)^2,
\ee
where the sum is taken over all particles in the system, which also includes all the resonances. 

While this calculation does not account for the dynamical effects of resonances, their widths are explicitly incorporated in the bulk viscosity. In that sense, the effect of long-lived resonances which potentially block the bulk relaxation are also included in the PHSD bulk viscosity calculation.

\item The Moroz calculation~\cite{Moroz:2013vd} uses the relaxation time approximation to analytically calculate the viscosities of the hadron gas in a similar fashion as to what was presented with the Chapman-Enskog formalism in Sec.~\ref{sec:bulk_pion}. In this framework, although all resonances are incorporated in the various cross-sections of the collision term, they do not per se exist as propagating particles in the calculation, and only binary elastic collisions are considered.

As this calculation is closer to our $\zeta_{{\rm eff}}/s$, we conjecture that the slow processes dominating $\zeta/s$ in SMASH are not included in the list of processes of~\cite{Moroz:2013vd}, or that the difference is due to dynamical effects not being included in that calculation.

\item  Let us comment now on the state-of-the-art values of $\zeta/s(T)$ extracted from hybrid models~\cite{Bernhard:2019bmu, Paquet:2020rxl,Everett:2020yty}. The temperature dependence follows some predefined ansatz, motivated by the Hagedorn picture of~\cite{NoronhaHostler:2008ju} where $\zeta/s$ increases with temperature. A Bayesian analysis is then employed to constrain the functional dependence using experimental data for bulk observables at RHIC and LHC energies. We show the final results for the temperature dependence of $\zeta/s$ at temperatures close to $T_c$ given in Ref.~\cite{Bernhard:2019bmu} in the 90\% credible region, and a more recent (yet unpublished) calculation given in~\cite{Paquet:2020rxl} in the 90\% confidence interval. The latter study uses 17 parameters (instead of 14 of the former) and applies closure tests to validate the Bayesian analyses. In addition, the parametrization of the prior for the bulk viscosity is different. The values of the bulk viscosity are of the same order as our $\zeta_{{\rm eff}}/s$ (which, we remind the reader, should at these high temperatures be considered a minimum value, since some contribution from the higher modes might be missing) but not compatible with $\zeta/s$. This is nicely consistent with the claim that in heavy-ion collisions, the slowest processes (whose inverse rates are larger than the fireball lifetime) do not play any role in the inferred bulk viscosity.
\end{itemize}

We finally comment on the comparison between two results of $\zeta/s$ above $T_c$. The first one~\cite{Karsch:2007jc} is based on lattice-QCD calculations, and the second one is obtained from Bayesian analyses using heavy-ion collisions~\cite{Bernhard:2019bmu}. While the first one shows a large enhancement of the bulk viscosity up to $\zeta/s \simeq 0.3$ near $T_c$, the second result shows a more smooth peak with a maximum of $\zeta/s \simeq 0.04$ (median value in green dotted line in our Fig.~\ref{bulk_comparison}). We argue that (part of) this discrepancy should come from the same conceptual difference between the inclusive and effective bulk viscosities. While the calculation in~\cite{Karsch:2007jc} is performed for a static QGP medium in equilibrium with all modes integrated, the method in Ref.~\cite{Bernhard:2019bmu} uses information from real heavy-ion collisions, and estimates the viscosity associated to an evolving, finite-lived system where slow modes do not have time to contribute. This difference becomes even more evident in the analysis of the chiral critical point in~\cite{Karsch:2007jc}, where a very slow critical mode provides the bulk viscosity with a critical exponent, making it divergent at $T_c$. In a dynamical system like a heavy-ion collision, these soft effects are not effective in practice, transforming any possible large increase of $\zeta/s$ into a milder peak like the one seen in~\cite{Bernhard:2019bmu}. It such a case the effective bulk viscosity is also blind to the very slow critical modes. This is another reason supporting the relevance of the distinction between an inclusive and an effective bulk viscosity in heavy-ion physics.

\section{Conclusions~\label{sec:conclusions}}

We have presented our estimates for the bulk viscosity and $\zeta/s$ of a hadron gas as a function of temperature between the range $T=80-170$ MeV at vanishing baryochemical potential. The results at the highest temperatures should be understood as a theoretical extrapolation, as the effects of a deconfined medium---which should take place at such temperatures---are not included.

The calculation of the bulk viscosity is numerically very challenging due to the small size of the fluctuations in the bulk channel, and because the statistical uncertainties in the pressure average can be of the same order. The systematic uncertainty is estimated by comparing to Chapman-Enskog calculations in a simple system with only one particle species. For the final results in a full hadron gas we can confirm that our calculation lies within the same area as previous calculations and extractions from experimental data in heavy-ion collisions. We observe a decreasing trend of $\zeta/s$ as a function of temperature, which needs to be reconciled with the expectation of a smooth maximum around the crossover transition to the quark-gluon plasma, which is absent in our model. 

We find that mass-changing processes, namely resonance excitations, have a very strong influence on the bulk viscosity. This is rather straightforward to understand since such processes allow to store kinetic energy in the mass of the particles and enhance the fluctuations of the kinetic energy of the system. Our results can be employed in future assumptions for the prior for Bayesian multi-parameter analyses and compared to lattice-QCD calculations once they become available.

One of our main results is the need for a distinction between the inclusive bulk viscosity $\zeta$ and the effective bulk viscosity $\zeta_{{\rm eff}}$. The first one is computed for long-lived systems in equilibrium, in which all components of the medium need to relax for the restoration to equilibrium to occur. We have explicitly shown that the slowest processes determine the bulk viscosity, as their contribution dominates the decay of the correlation function. These modes with relaxation times of several dozens and even hundreds of fm/$c$ make the $\zeta/s$ a large coefficient for all temperatures. 

The effective bulk viscosity is the coefficient controlling the relaxation to equilibrium of systems with a finite lifetime, as the ones happening in RHICs. The very slow modes do not have enough time to play a role and their contribution to the correlation function are explicitly removed. This $\zeta_{{\rm eff}}/s$, rather than the inclusive $\zeta/s$, seems to be the analogue to the extracted bulk viscosity over entropy density in relativistic hydrodynamic simulations, as such computations pretend to describe the realistic real-time evolution of RHICs. The state-of-the-art simulations, implementing Bayesian techniques to extract this (and other) parameters~\cite{Bernhard:2016tnd, Bernhard:2019bmu,Paquet:2020rxl,Everett:2020yty} provide values of $\zeta/s \simeq 0.04-0.1$ for temperatures between $T=140$ MeV and $T=160$ MeV. Our effective bulk viscosity is compatible with these numbers, whereas the inclusive $\zeta/s$ is not consistent with them. Such a difference illustrates once more the need of both kind of coefficients. The Green-Kubo method applied to a static box in equilibrium allows the system to wait until all microscopic processes have relaxed---including the very slow ones---and incorporate all of them to the ``genuine'' transport coefficient of the hadron gas. However, in heavy-ion collisions these very slow modes cannot keep up with the dynamics of the fireball and remain out of equilibrium until freeze-out occurs. Any fast process able to relax during the fireball lifetime, will have an effect into the transport coefficients, and eventually, into the experimental observables which depends on those (flow harmonics, mean transverse momentum...). On the contrary, the inferred transport coefficients via hydrodynamic simulations of RHICs should rarely be affected by the slow modes. 

The effect of these very slow modes was not observed in other coefficients like the shear viscosity, electrical conductivity or cross-conductivities~\cite{Rose:2017bjz,Hammelmann:2018ath,Rose:2020sjv}, where the single exponential decay was found to be a good approximation (except for very high temperatures where the system becomes dense). This situation illustrates that the bulk viscosity is a much more subtle quantity than other transport coefficients; being extremely dependent on the microscopical details of the interactions, any comparison between calculations must be performed with caution.

\begin{acknowledgments}

Computational resources have been provided by the Center for Scientific Computing (CSC) at the Goethe-University of Frankfurt. Funded by the Deutsche Forschungsgemeinschaft (DFG, German Research Foundation) – Project number 315477589 – TRR 211. J.M.T.-R. also acknowledges partial support from DFG Project number 411563442. This work was supported by the Helmholtz International Center for the Facility for Antiproton and Ion Research (HIC for FAIR) within the
framework of the Landes-Offensive zur Entwicklung Wissenschaftlich-Oekonomischer Exzellenz (LOEWE) program launched by the State of Hesse.

\end{acknowledgments}

\appendix

\section{Correlation function at $t=0$ and other thermodynamical quantities~\label{app:correl}}

In Sec.~\ref{sec:calibration} we obtained the expression for the bulk correlation function at $t=0$,
\be \label{eq:C0bis}
C_\zeta(0) = \frac{g}{V} \int \frac{d^3p}{(2\pi)^3} \frac{1}{E_p^2} \left[ \frac{p^2}{3} - E_p^2  \left( \frac{\partial P}{\partial \epsilon} \right)_n  -E_p  \left( \frac{\partial P}{\partial n} \right)_\epsilon  \right]^2 \exp\left(- \frac{E_p-\mu}{T}\right) \ , \ee
with $E_p=\sqrt{p^2+m^2}$, and where we have written explicitly the Boltzmann distribution $f^{{\rm eq}}(p)=g \exp[-(E_p-\mu)/T]$. This quantity is a function of temperature and chemical potential. For the pion gas with only elastic interactions, we take the pion pseudochemical potential to zero without loss of generality, while for the full hadron gas the baryochemical potential is set to zero in this work.

The value of $C_\zeta(0)$ depends itself on other thermodynamical quantities. In the ideal gas limit, these can be expressed in terms of some integrals ${\cal J}_{n,k}(T,\mu)$, as done in Ref.~\cite{Torres-Rincon:2012sda} but for a Boltzmann gas in the local rest frame,
\be \label{eq:Jfuncs} {\cal J}_{n,k}(T,\mu) = \frac{g}{(2k+1)!!} \int \frac{d^3p}{(2\pi)^3} p^{2k} E_p ^{n-2k-1}  \ \exp\left(- \frac{E_p-\mu}{T}\right)  \ . \ee

The particle and entropy densities can be expressed as
\be n = {\cal J}_{2,1}/T \ , \quad s= ( {\cal J}_{3,1} -\mu {\cal J}_{2,1})/T^2 \ . \ee
For the quantities used in Eq.~(\ref{eq:C0bis}) the relations become much more complicated. To simplify the expressions, let us consider the case $\mu=0$ from now on, which is the one taken in this work. We obtain
\begin{align} 
 \left( \frac{\partial P}{\partial \epsilon} \right)_n & = \frac{{\cal J}_{3,1} {\cal J}_{1,0} - {\cal J}_{2,1} {\cal J}_{2,0} }{{\cal J}_{3,0}{\cal J}_{1,0}-{\cal J}_{2,0}^2} \ , \label{eq:vn} \\
 \left( \frac{\partial P}{\partial n} \right)_\epsilon & =  \frac{{\cal J}_{2,1} {\cal J}_{3,0} - {\cal J}_{3,1} {\cal J}_{2,0} }{{\cal J}_{3,0}{\cal J}_{1,0}-{\cal J}_{2,0}^2}  \ . \label{eq:kappa}
 \end{align}
 
The adiabatic speed of sound (\ref{eq:vs}) reads
\be v_S^2  =  \left( \frac{\partial P}{\partial \epsilon} \right)_n + \frac{n}{sT} \left( \frac{\partial P}{\partial n} \right)_\epsilon = \frac{{\cal J}_{3,1}^2 {\cal J}_{1,0}-2 {\cal J}_{2,1} {\cal J}_{2,0} {\cal J}_{3,1} + {\cal J}_{2,1}^2 {\cal J}_{3,0}}{{\cal J}_{3,1}({\cal J}_{3,0}{\cal J}_{1,0}-{\cal J}_{2,0}^2)} \ . \label{eq:vs2} \ee

In a similar fashion $C_\zeta(0)$ can be expressed in terms of ${\cal J}_{n,k}(T,\mu)$ functions if desired.

We plot some of these thermodynamic quantities as functions of $T$ for different systems containing several hadron species. We consider $\pi,K,N,\rho,K^*,\Delta$, where for the resonances we need to generalize the expression~(\ref{eq:Jfuncs}) to include an additional integral over their spectral functions.

\begin{figure}[ht]
\centering
\includegraphics[width=65mm]{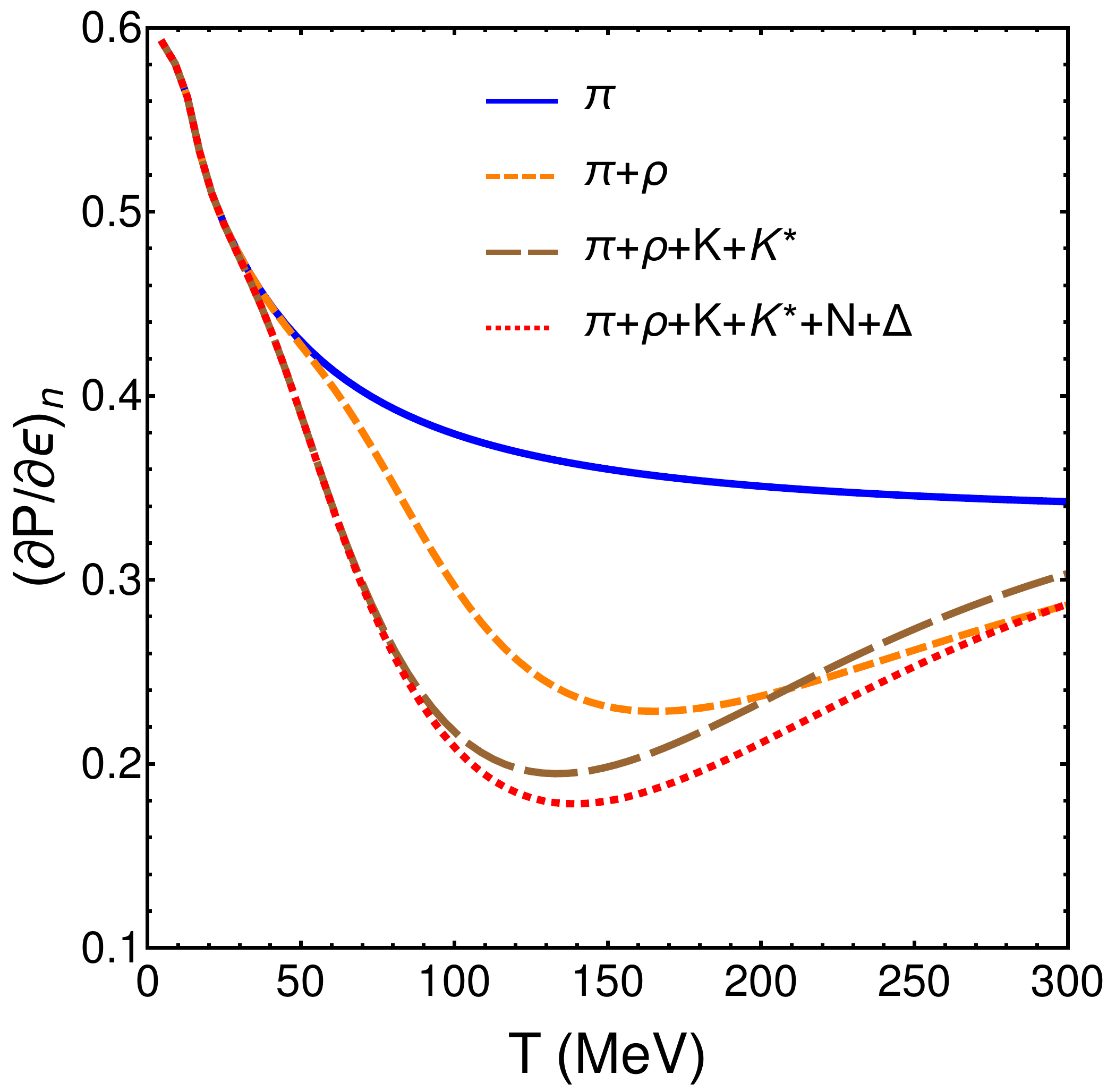}
\includegraphics[width=70mm]{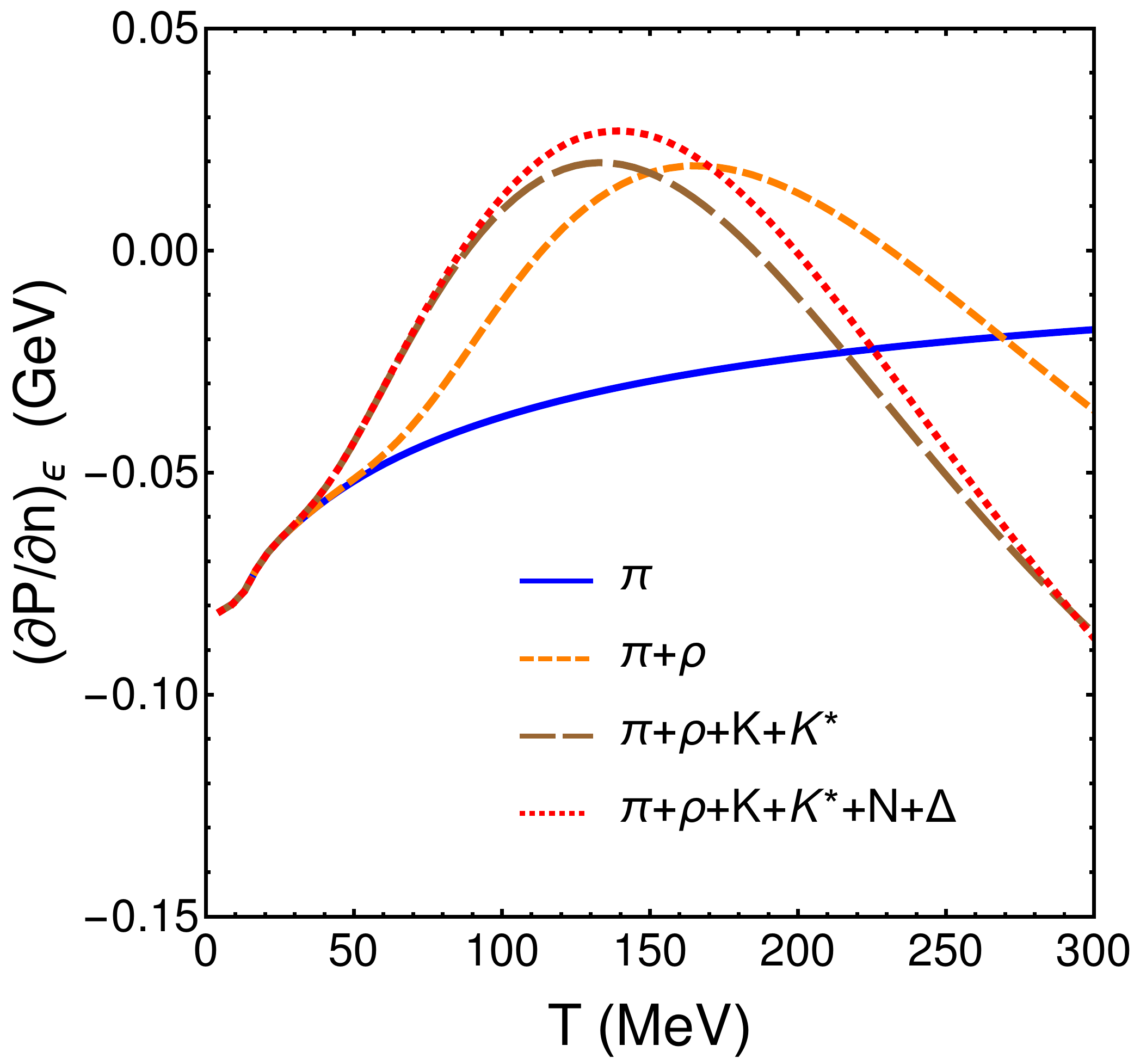}
\includegraphics[width=72mm]{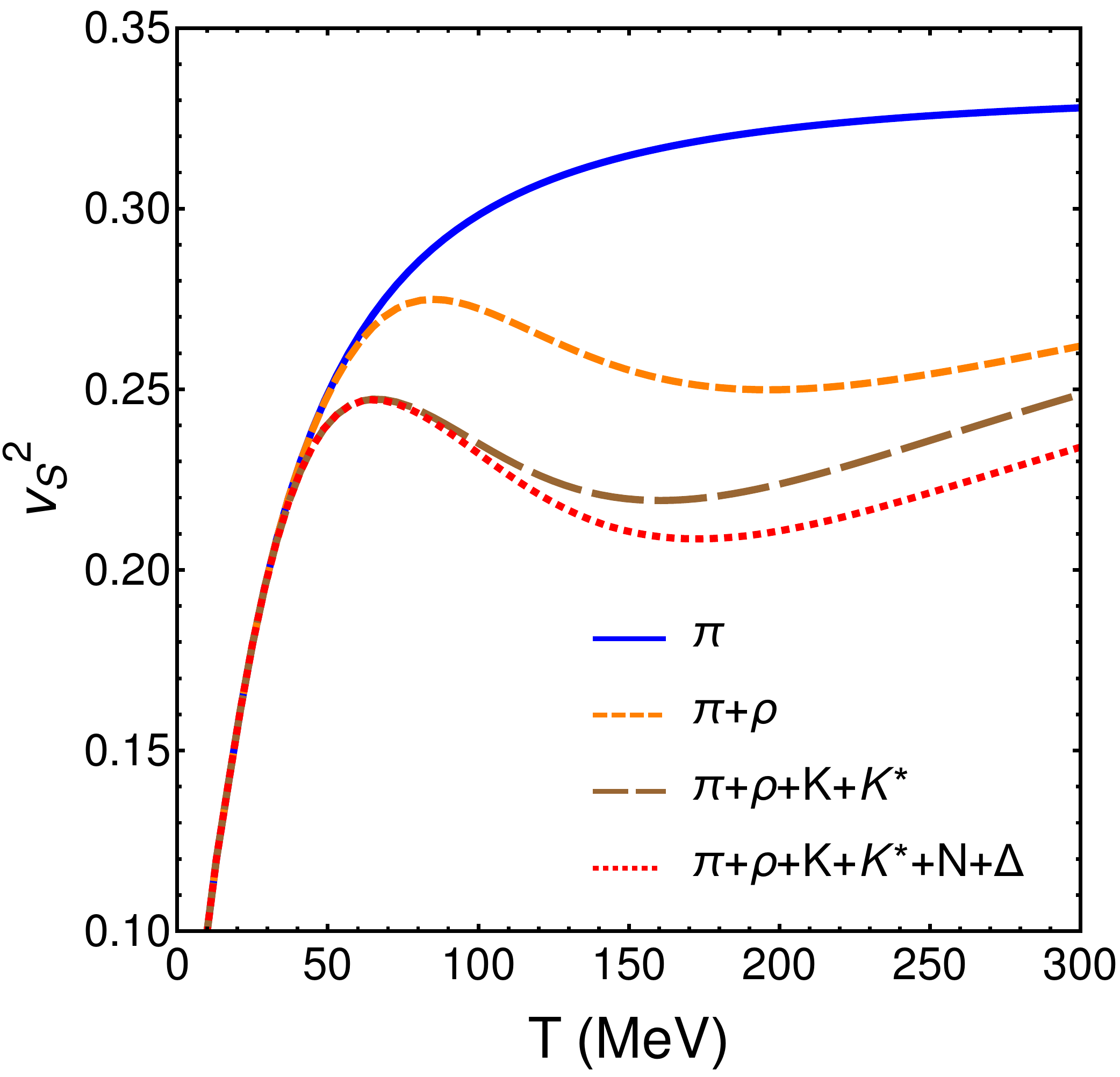}
\caption{Thermodynamic quantities (\ref{eq:vn}), (\ref{eq:kappa}) and (\ref{eq:vs2}) versus the temperature as a function of hadron components in an ideal hadron gas.}
\label{fig:thermo}
\end{figure}

In Fig.~\ref{fig:thermo} we present the quantities (\ref{eq:vn}), (\ref{eq:kappa}) and the adiabatic speed of sound (\ref{eq:vs2}), for a hadron gas when several species are subsequently introduced. Notice that $v_S^2$ already presents a nonmonotonous behavior when $\rho$ mesons are introduced in the pion gas. For a more realistic case with more states covering higher masses, we refer to Fig.~\ref{fig:speedsound}.

Let us finally comment on two particular cases which can be quite illustrative, although they are not used in the results of this paper.
For massless particles, we note that
\be {\cal J}_{n,k}(T) = \frac{gT^{n+2}}{2\pi^2} \frac{\Gamma(n+2)}{(2k+1)!!}\ , \ee
and one obtains $\left( \frac{\partial P}{\partial n} \right)_\epsilon=0$, $ \left( \frac{\partial P}{\partial \epsilon} \right)_n= \left( \frac{\partial P}{\partial \epsilon} \right)_{S}=1/3$, so the bulk viscosity is seen to vanish proportionally to the square of $1/3-v_S^2$~\cite{Weinberg:1972kfs,Arnold:2006fz}. 

For an ensemble where the particle number $n$ is not conserved, one does not introduce any chemical potential, and the thermodynamic functions only depends on $T$. The speed of sound reduces to,
\be v^2_S = \left( \frac{\partial P}{\partial \epsilon} \right)_S = \frac{dP/dT}{d\epsilon/dT} = \frac{{\cal J}_{3,1}}{{\cal J}_{3,0}} \ . \ee

The $C_\zeta(0)$ in this particular case would read
\be
C_\zeta(0)= \frac{g}{V} \int \frac{d^3p}{(2\pi)^3} \frac{1}{E_p^2} \left( \frac{p^2}{3} - E_p^2  v^2_S    \right)^2 \ \exp\left(- \frac{E_p}{T}\right) \ . 
\ee
which can be further simplified to
\be \label{eq:C0simp}
C_\zeta(0)= \frac{g}{V} \int \frac{d^3p}{(2\pi)^3}  \left[ \left( \frac13 - v^2_S \right) E_p - \frac{m^2}{3E_p}  \right]^2 \ \exp\left(- \frac{E_p}{T}\right) \ . 
\ee

Combining the expression for the bulk viscosity in Eq.~(\ref{final_bulk_eq}) and our previous result on the shear viscosity~\cite{Rose:2017bjz} (also using the exponential decay ansatz) we obtain
\be \frac{\zeta}{\eta} = \frac{C_\zeta(0) \tau_\zeta}{C_\eta(0) \tau_\eta} \ , \ee
where
\be C_\eta (0)= \frac{1}{15V}\int \frac{d^3p}{(2\pi)^3} f^{{\rm eq}}(p) \frac{p^4}{E^2_p} \ , \ee
is the shear correlation function at $t=0$, and $\tau_\eta$ is the relaxation time of a fluctuation in the shear channel.
If we assume that $\tau_\eta \simeq \tau_\zeta$ and introduce the result~(\ref{eq:C0simp}) for massless particles ($E_p = p$) one gets
\be \frac{\zeta}{\eta} \simeq \frac{1}{V} \int \frac{d^3p}{(2\pi)^3}   \left( \frac13 - v^2_S \right)^2 p^2 f^{{\rm eq}} (p) \Bigg/ \frac{1}{15V}\int \frac{d^3p}{(2\pi)^3} p^2 f^{{\rm eq}}(p) =  15 \left( \frac13-v_S^2\right)^2 \ , \ee
which coincides with the well-known relation~(\ref{eq:conformality}) also in the numerical factor.

\section{Multi-exponential fitting~\label{app:fitting}}

We fit the correlation functions given in Fig.~\ref{fig:bulk_correls_full} to the form
 \be C_\zeta(t) \simeq \sum_{i=1}^3 C_{\zeta,i} (0) \exp \left( -t/\tau_{\zeta,i} \right)\ , \label{eq:lincomexp3}\ee 
using different methods. First of all, to check that all modes are indeed exponential, we proceed with a sequential method, as described at the end of Sec.~\ref{sec:full_bulk_section}: one finds the exponential fit of the tail of $C_\zeta(t)$ and then subtracts the fitted component from the full correlation function. Then, one repeats the procedure to find the exponential decay of the intermediate range of times, and after another subtraction, one fits the small-$t$ part of the function.

\begin{figure}[ht]
  \centering
  \includegraphics[width=80mm]{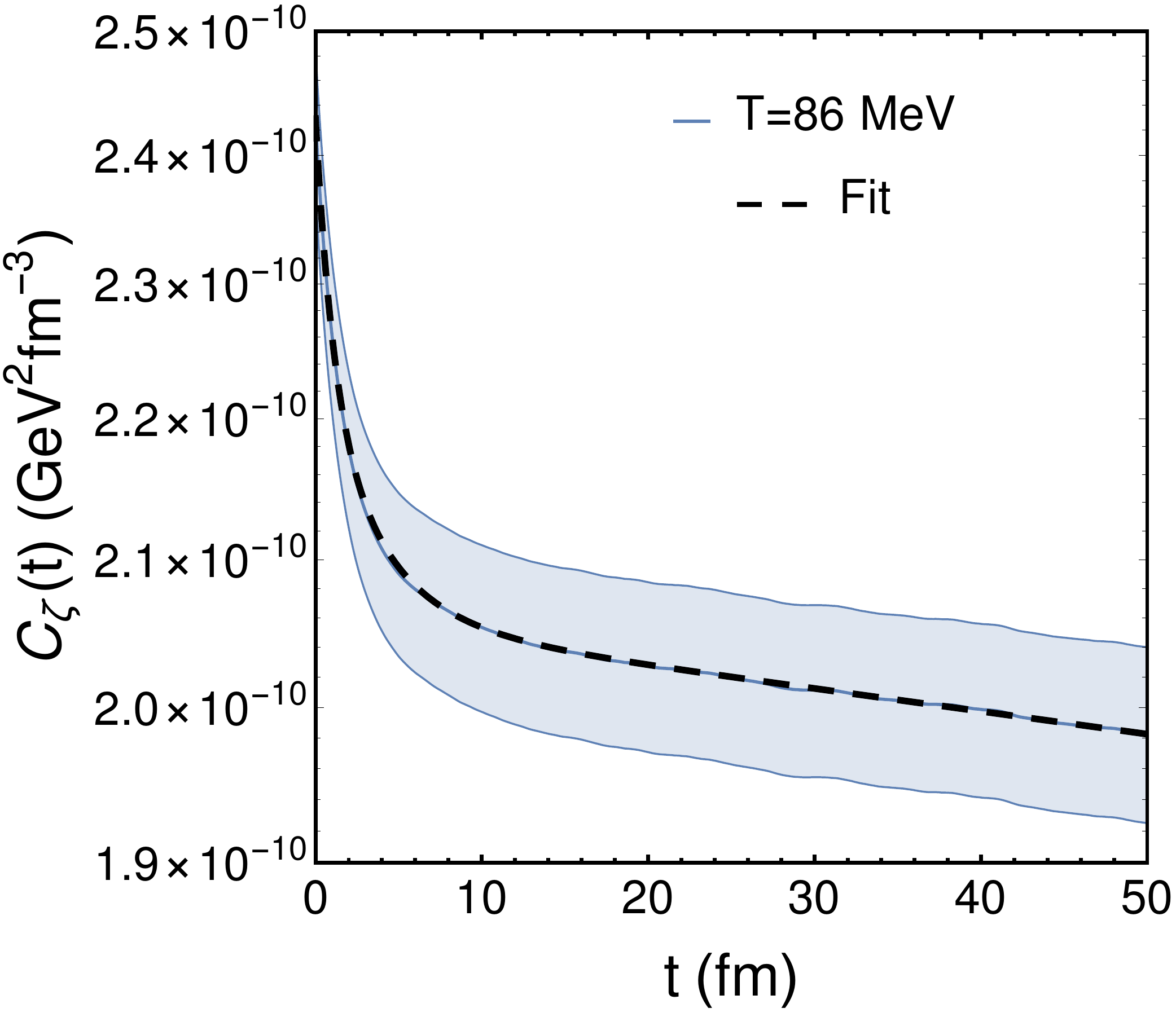}
\caption{Example of the triple exponential fit to the correlation function obtained from SMASH for the full hadron gas at $T=86$ MeV. We show the resulting sequential fit in dashed line on top of the error band of $C(t)$.}
  \label{fig:bulk_correls}
\end{figure}

We present an example of such a fit in Fig.~\ref{fig:bulk_correls} for the temperature of $T=86$ MeV (the one with largest error bars). It is difficult to assign an uncertainty to the sequential fit itself, due to the rather manual procedure, so it is given as is.

The quality of the fit is very good. We double-check the resulting fit procedure against a global fit of $C_\zeta(t)$ using the {\it NonlinearModelFit} option in Mathematica~\cite{Mathematica}, and also see that using a larger number $N$ of exponentials results in a poorer quality of the fit, as some components have negative amplitudes, which is physically unreasonable.

\begin{table}[ht]
\begin{center}
 \begin{tabular}{|c||cc|cc|cc|} 
 \hline
 \multicolumn{7}{|c|}{Sequential fit} \\
 \hline
 $T$ (MeV) & $C_{\zeta,1} (0)$ (GeV$^2$fm$^{-3}$) & $\tau_{\zeta,1}$ (fm) &$C_{\zeta,2} (0)$ (GeV$^2$fm$^{-3}$)& $\tau_{\zeta,2}$ (fm) &$C_{\zeta,3} (0)$ (GeV$^2$fm$^{-3}$) & $\tau_{\zeta,3}$ (fm) \\ 
\hline
86 & 1.96$\cdot 10^{-11}$ & 1.00 & 1.74$\cdot 10^{-11}$ & 3.55 & 2.06$\cdot 10^{-10}$ & 1328.01 \\
114 & 1.01$\cdot 10^{-9}$ & 1.16 & 2.44$\cdot 10^{-10}$ & 5.47 & 1.28$\cdot 10^{-9}$ & 92.80 \\
142  & 1.03$\cdot 10^{-8}$ & 0.66 & 6.39$\cdot 10^{-9}$ & 2.65 & 1.75$\cdot 10^{-8}$ & 34.81 \\
172  & 2.09$\cdot 10^{-7}$ & 0.42 & 1.85$\cdot 10^{-7}$ & 1.77 & 5.40$\cdot 10^{-7}$ & 19.31 \\
\hline
 \multicolumn{7}{|c|}{Global fit} \\
\hline
86 & ($2.49 \pm 1.61$)$\cdot 10^{-11}$ & $1.21\pm 0.53$ & ($1.23\pm 0.69$)$\cdot 10^{-11}$ & $4.3\pm2.7$ & ($2.05\pm0.01$)$\cdot 10^{-10}$ & $1388\pm262$ \\
114 & ($9.59\pm0.80$)$\cdot 10^{-10}$ & $1.10\pm0.10$ & ($2.97\pm1.99$)$\cdot 10^{-10}$ & $4.9\pm1.5$ & $(1.28\pm0.03)$ $\cdot 10^{-9}$ & $93.0\pm6.8$ \\
142  & ($1.15\pm0.09$)$\cdot 10^{-8}$ & $0.73\pm0.06$ & ($5.52\pm0.69$)$\cdot 10^{-9}$ & $3.3\pm0.6$ & ($1.71\pm0.03$)$\cdot 10^{-8}$ & $36.1\pm1.3$ \\
172  & ($2.86\pm0.15$)$\cdot 10^{-7}$ & $0.58\pm0.05$ & ($1.50\pm0.09$)$\cdot 10^{-7}$ & $3.7\pm0.7$ & ($4.91\pm0.17$)$\cdot 10^{-7}$ & $21.6\pm0.7$ \\
\hline
\end{tabular}
\caption{Results from the fits with 3 modes. Above: Component-to-component fit, with individual sequential subtractions. Below: Fit to ansatz in one global fit using the ROOT library.
}
\label{tab:fits}
\end{center}
\end{table}

The parameters of the ``sequential fits'' for all temperatures are summarized in Table~\ref{tab:fits} in the upper block of data. All fits have been checked against independent fits in Mathematica (not shown here). 

We apply yet another method by making a global fit using ROOT~\cite{Brun:1997pa}, which takes into account the error band of $C_\zeta(t)$ and also provides the statistical uncertainties of the fitting parameters. The outcome of these fits is shown in the lower block of data of Table~\ref{tab:fits}. The numbers are more or less consistent with the ``sequential fit'', although some deviations remain. Notice that the sequential fit carries an additional systematic error (also difficult to extract) coming from the selection of fit ranges, which has to be decided relatively arbitrarily. In every case, we checked that both independent fits in Table~\ref{tab:fits} describe the correlation function really well, and they actually result in a very similar bulk viscosity. In the main text, the global fit by ROOT is used because it provides a measure of its statistical uncertainty.

For completeness we also provide the results for the global fit in the case of the ``smaller'' boxes for the same full hadron system. They are shown in Table~\ref{tab:fits2} only for the case of the fits using ROOT package.

\begin{table}[ht]
\begin{center}
 \begin{tabular}{|c||cc|cc|cc|} 
  \hline
  \multicolumn{7}{|c|}{Global fit} \\
\hline
 $T$ (MeV) & $C_{\zeta,1} (0)$ (GeV$^2$fm$^{-3}$) & $\tau_{\zeta,1}$ (fm) &$C_{\zeta,2} (0)$ (GeV$^2$fm$^{-3}$)& $\tau_{\zeta,2}$ (fm) &$C_{\zeta,3} (0)$ (GeV$^2$fm$^{-3}$)& $\tau_{\zeta,3}$ (fm) \\ 
 \hline
86 & ($1.35 \pm 0.18$)$\cdot 10^{-10}$ & $1.36\pm 0.26$ & ($4.08\pm 1.51$)$\cdot 10^{-11}$ & $7.2\pm4.2$ & ($3.77\pm0.07$)$\cdot 10^{-10}$ & $616\pm171$ \\
114 & ($2.97\pm0.33$)$\cdot 10^{-9}$ & $1.02\pm0.11$ & ($1.26\pm0.30$)$\cdot 10^{-9}$ & $4.0\pm1.0$ & $(4.03\pm0.07)$$\cdot 10^{-9}$ & $84.4\pm4.3$ \\
142  & ($9.63\pm0.72$)$\cdot 10^{-8}$ & $0.77\pm0.06$ & ($3.62\pm0.51$)$\cdot 10^{-8}$ & $3.6\pm1.0$ & ($1.38\pm0.04$)$\cdot 10^{-7}$ & $32.9\pm1.3$ \\
172  & ($2.28\pm0.06$)$\cdot 10^{-6}$ & $0.57\pm0.03$ & ($1.98\pm0.19$)$\cdot 10^{-6}$ & $5.6\pm0.7$ & ($3.05\pm0.24$)$\cdot 10^{-6}$ & $25.7\pm1.9$ \\
\hline
\end{tabular}
\caption{Results from the fits with 3 modes for the ``smaller box''. The fit to the ansatz (\ref{eq:lincomexp3}) is done with the ROOT library.
}
\label{tab:fits2}
\end{center}
\end{table}

\bibliographystyle{ieeetr}
\bibliography{bulk_paper}

\end{document}